\documentclass[english,british,refpage,intoc,bibliography=totoc,index=totoc,BCOR7.5mm,captions=tableheading]{amsart}
\usepackage{lmodern}

\usepackage[T1]{fontenc}
\usepackage[latin9]{inputenc}
\usepackage{geometry}
\usepackage{xcolor}
\usepackage{babel}
\usepackage{verbatim}
\usepackage{units}
\usepackage{enumitem}
\usepackage{amsbsy}
\usepackage{amstext}
\usepackage{amsthm}
\usepackage{amssymb}
\usepackage[unicode=true,
 bookmarks=true,bookmarksnumbered=true,bookmarksopen=false,
 breaklinks=false,pdfborder={0 0 1},backref=false,colorlinks=true]
 {hyperref}
\hypersetup{pdftitle={The LyX User's Guide},
 pdfauthor={LyX Team},
 pdfsubject={LyX},
 pdfkeywords={LyX},
 linkcolor=black, citecolor=black, urlcolor=blue, filecolor=blue, pdfpagelayout=OneColumn, pdfnewwindow=true, pdfstartview=XYZ, plainpages=false}
\usepackage{breakurl}

\makeatletter

\providecommand{\tabularnewline}{\\}

\numberwithin{equation}{section}
\numberwithin{figure}{section}
\theoremstyle{plain}
\newtheorem{thm}{\protect\theoremname}[section]
  \theoremstyle{remark}
  \newtheorem{rem}[thm]{\protect\remarkname}
  \theoremstyle{plain}
  \newtheorem{lem}[thm]{\protect\lemmaname}
  \theoremstyle{definition}
  \newtheorem{example}[thm]{\protect\examplename}

\usepackage{upgreek}
\usepackage{amssymb}
\usepackage{mathtools}
\let\originalleft\left
\let\originalright\right
\renewcommand{\left}{\mathopen{}\mathclose\bgroup\originalleft}
\renewcommand{\right}{\aftergroup\egroup\originalright}

\makeatother

  \addto\captionsbritish{\renewcommand{\examplename}{Example}}
  \addto\captionsbritish{\renewcommand{\lemmaname}{Lemma}}
  \addto\captionsbritish{\renewcommand{\remarkname}{Remark}}
  \addto\captionsbritish{\renewcommand{\theoremname}{Theorem}}
  \addto\captionsenglish{\renewcommand{\examplename}{Example}}
  \addto\captionsenglish{\renewcommand{\lemmaname}{Lemma}}
  \addto\captionsenglish{\renewcommand{\remarkname}{Remark}}
  \addto\captionsenglish{\renewcommand{\theoremname}{Theorem}}
  \providecommand{\examplename}{Example}
  \providecommand{\lemmaname}{Lemma}
  \providecommand{\remarkname}{Remark}
\providecommand{\theoremname}{Theorem}

\begin{document}

\title[Augmented Dimensional Analysis]{Theory and Application of Augmented Dimensional Analysis }

\author{Dan Jonsson}
\begin{abstract}
We present an innovative approach to dimensional analysis, referred
to as augmented dimensional analysis and based on a representation
theorem for complete quantity functions with a scaling-covariant scalar
representation. This new theorem, grounded in a purely algebraic theory
of quantity spaces, allows the classical $\pi$ theorem to be restated
in an explicit and precise form and its prerequisites to be clarified
and relaxed. Augmented dimensional analysis, in contrast to classical
dimensional analysis, is guaranteed to take into account all relations
among the quantities involved. Several examples are given to show
that the information thus gained, together with symmetry assumptions,
can lead to new or stronger results. We also explore the connection
between dimensional analysis and matroid theory, elucidating combinatorial
aspects of dimensional analysis. It is emphasized that dimensional
analysis rests on a principle of covariance.
\end{abstract}

\maketitle

\section{Introduction\label{sec:I}}

In classical dimensional analysis, the main result is the so-called
$\pi$ theorem, with a long history featuring contributions by Fourier
\cite{fo}, Vaschy \cite{va}, Federman \cite{fe}, Buckingham \cite{bu}
and others. The $\pi$ theorem helps to transform a ``physically
meaningful'' equation
\begin{equation}
t_{0}=\phi\left(t_{1},\ldots,t_{n}\right),\label{eq:pi1}
\end{equation}
expressing a relationship among quantities, into a more informative
equation, using data about the dimensions of $t_{0},\ldots,t_{n}$.
This can be done by representing the unknown function $\phi$ as a
product of the form $\prod_{j=1}^{r}\nolimits\!x{}_{j}^{K_{j}}\,\psi$,
where $\psi$, while also unspecified, is a function of fewer arguments
than $\phi$. Specifically, for a suitable partition $\left\{ \left\{ x_{1},\ldots,x_{r}\right\} ,\left\{ y_{1},\ldots,y_{n-r}\right\} \right\} $
of $\left\{ t_{1},\ldots,t_{n}\right\} $,
\begin{equation}
t_{0}=\prod_{j=1}^{r}\nolimits\!x{}_{j}^{K_{j}}\,\psi\left(\pi_{1},\ldots,\pi_{n-r}\right)\qquad\left(1\leq r\leq n\right),\label{pi2}
\end{equation}
where $\pi_{i}$ is a ``dimensionless product'' of the form
\begin{equation}
\pi_{i}=y_{i}\prod_{j=1}^{r}\nolimits\!x{}_{j}^{-K_{ij}}.\label{pi3}
\end{equation}

In classical dimensional analysis, $\phi$ is a real function, and
$t_{0},\ldots,t_{n}$ are measures of physical quanti\-ties. Various
assumptions pertaining to the functions $\phi$ and $\psi$, the variables
$t_{i}$ and the parameters $K_{j},K_{ij}$ have been made in connection
with the development of the theory of dimensional analysis.
\begin{enumerate}
\item Proofs of the $\pi$ theorem rely on assumptions that $\phi$, and
hence $\psi,$ have certain mathematical properties. Generally speaking,
these assumptions fall into two groups:
\begin{enumerate}
\item Those that translate the fact that $\phi$ is ``physically meaningful''
into mathematical terms, for example, the assumption proposed by Langhaar
\cite{la} and Brand \cite{bra} that $\phi$ satisfies a generalized
homogeneity condition.
\item Those that are made for internal mathematical reasons. For example,
proofs of the $\pi$ theorem have involved assumptions that $\phi$
is continuous or differentiable or analytic \cite[pp. 67--71]{gi};
such assumptions may relate more to availability of mathematical notions
and techniques than to what is ``physically meaningful''.
\end{enumerate}
\item It is usually assumed that $t_{i}>0$. Some authors consider quantities
to be positive by nature, but there are also formal reasons for this
assumption: it is required in some proofs of the $\pi$ theorem. This
restriction is not essential, however.
\item The exponents $K_{j}$ and $K_{ij}$ are usually assumed to be rational
or real numbers \cite[p. 293]{ca}, but some authors use only integer
exponents \cite{fle,qua,ra2}.
\item It is usually implicitly assumed that for any $\phi$ there is just
one $\psi$ such that (\ref{pi2}) holds, or at least that it suffices
to consider one $\psi$, because supposedly all $\psi$ are equivalent
or all but one $\psi$ can be disregarded. This way of thinking is
challenged in \cite{jo1,key-6} (see also \cite{whi,key-11}).
\end{enumerate}
The emergence of quantity calculus \cite{bo}, and more recent related
developments \cite{dro,fle,qua,whi-1,ca,sze,key-2,key-1,jo1,ra1,jo3},
have led to attempts to formulate versions of the $\pi$ theorem where
$t_{0},\ldots,t_{n}$ are the quantities measured rather than their
numerical measures \cite{dro,qua,whi-1,jo1,ra2,har,jo2}. The present
contribution belongs to this quantity calculus tradition. Rather than
real functions $\phi$ and $\psi$ we use quantity functions $\Phi$
and $\Psi$ on a quantity space $Q$ over a field $\mathbb{K}$ (see
Section \ref{sec:2}). In this new context, too, the aspects considered
in (1)\textendash (4) need to be addressed. It turns out that it is
possible to extend the scope of classical dimensional analysis by
modifying and clarifying its foundations.
\begin{enumerate}
\item[\textcolor{blue}{(i)}] The idea that $\Phi$ is a ``physically meaningful'' quantity function
is expressed by the purely alge\-braic requirement that $\Phi$ is
a quantity function that is algebraically equivalent to a ``complete''
quantity function $\Phi^{*}$ with a ``covariant scalar representation''
(see Section \ref{sec:2}). 
\item[\textcolor{blue}{(ii)}] $Q$ can be a quantity space over any field, so the (generalized)
measures of the quantities $t_{0},\ldots,t_{n}$ need not be real
numbers. The quantities $x_{1},\ldots,x_{r}$ in (\ref{pi2}) and
(\ref{pi3}) are assumed to be non-zero, but there are no further
absolute constraints on $t_{0},\ldots,t_{n}$ or their measures (see
the discussion in connection with Lemma \ref{t1}).
\item[\textcolor{blue}{(iii)}] Only integers are used in exponents, so we write $y^{2}=x$ instead
of $y=x^{\frac{1}{2}}$ etc. Specifically, in (\ref{eq:pi1}) and
(\ref{pi2}) we replace $t_{0}$ by $t_{0}^{\mathfrak{K}}$, where
$\mathfrak{K}$ is a positive integer, in (\ref{pi2}) $K_{j}$ are
integers, and in (\ref{pi3}) $K_{ij}$ are integers and $y_{i}$
is replaced by $y_{i}^{K_{i}}$, where $K_{i}$ are positive integers.
\item[(iv)] There may be more than one partition $\left\{ \left\{ x_{1},\ldots,x_{r}\right\} ,\left\{ y_{1},\ldots,y_{n-r}\right\} \right\} $
of $\left\{ t_{1},\ldots,t_{n}\right\} $ such that there is a quantity
function $\Psi$ and $\mathfrak{K}$, $K_{j}$, $K_{i}$, $K_{ij}$
such that $t_{0}^{\mathfrak{K}}=\prod_{j=1}^{r}\nolimits x{}_{j}^{K_{j}}\,\Psi\left(\pi_{1},\ldots,\pi{}_{n-r}\right)$,
where $\pi_{i}=y_{i}^{K_{i}}\prod_{j=1}^{r}\!x_{j}^{-K_{ij}}$. Thus,
the augmented dimensional analysis developed here does not yield a
single equation as in the classical approach but a system of $\Theta\geq1$
equations that should be considered collectively, 
\[
\begin{cases}
t_{0}^{\mathfrak{K}}=\prod_{j=1}^{r}\nolimits x{}_{1,j}^{K_{1,j}}\,\Psi_{1}\left(\pi_{1,1},\ldots,\pi{}_{1,n-r}\right),\\
\cdots\\
t_{0}^{\mathfrak{K}}=\prod_{j=1}^{r}\nolimits x{}_{\Theta,j}^{K_{\Theta,j}}\,\Psi_{\Theta}\left(\pi_{\Theta,1},\ldots,\pi{}_{\Theta,n-r}\right).
\end{cases}
\]
\end{enumerate}
Section 2, following this Introduction, is devoted to preliminaries.
The main representation theorems are stated and proved in Section
3, and their use in dimensional analysis is described in Section 4.
Several examples of dimensional analysis are given in Section 5, Section
6 connects dimensional analysis to matroid theory, and Section 7 looks
at dimensional analysis from the point of view of covariance.
\begin{rem}
In this article, we mainly describe an ``unbalanced'' approach to
dimensional analysis, where a dependent variable is chosen at the
outset. In Section \ref{sec:6-1}, however, we will also consider
a ``balanced'' approach, where the choice of dependent variable
is suspended.
\end{rem}

\section{\textcolor{blue}{Preliminaries\label{sec:2}}}

\subsection{\label{subsec:21}Quantity spaces and free abelian groups of dimensions}

In this article, we we will use the formalization of quantities and
dimensions presented and developed in some detail in \cite{jo3}.
For the sake of completeness, we briefly review some definitions and
results from \cite{jo3} here. 

A \emph{scalable monoid} over a ring $R$ is a monoid $Q$ equipped
with an $R$-action $\cdot$ on $Q$, 
\[
R\times Q\rightarrow Q,\qquad\left(\alpha,q\right)\mapsto\alpha\cdot q,
\]
compatible with multiplication in $Q$. For any $\alpha,\beta\in R$
and $q,q'\in Q$, we have $1\cdot q=q$, $\alpha\cdot\left(\beta\cdot q\right)=\alpha\beta\cdot q$
and $\alpha\cdot qq'=\left(\alpha\cdot q\right)q'=q\left(\alpha\cdot q'\right)$;
as a consequence, $\left(\alpha\cdot q\right)\left(\beta\cdot q'\right)=\alpha\beta\cdot qq'$.
We denote the identity element of $Q$ by $1_{\!Q}$, and set $q^{0}=1_{\!Q}$
for any $q\in Q$. An element $q\in Q$ may have an inverse $q^{-1}\in Q$
such that $qq^{-1}=q^{-1}q=1_{\!Q}$.

A \emph{finite} \emph{(quantity-space) basis} for a commutative scalable
monoid $Q$ over a field $\mathbb{K}$ is a set $E_{Q}=\left\{ e_{1},\ldots,e_{m}\right\} $
of invertible elements of $Q$ such that every $q\in Q$ has a unique
expansion 
\[
q=\mu_{E_{Q}}\left(q\right)\cdot\prod_{j=1}^{m}\nolimits e_{j}^{K_{j}},
\]
where $\mu_{E_{Q}}\left(q\right)\in\mathbb{K}$ and $K_{j}\in\mathbb{Z}$.
A \emph{(finitely generated) quantity space} $Q$ is a commutative
scalable monoid over a field, such that there exists a finite basis
for $Q$. The elements of a quantity space are called \emph{quantities}.
We may think of $\mu_{E_{Q}}\left(q\right)$ as the measure of $q$
relative to the derived unit $\prod_{j=1}^{m}\nolimits e_{j}^{K_{j}}$
in a coherent system of units, and indirectly relative to the base
units in $E_{Q}$.

The relation $\sim$ on $Q$ defined by $q\sim q'$ if and only if
$\alpha\cdot q=\beta\cdot q'$ for some $\alpha,\beta\in\mathbb{K}$
is a congruence on $Q$. The corresponding equivalence classes are
called \emph{dimensions}; $\left[q\right]$ is the dimension that
contains $q$. We have $\left[\lambda\cdot q\right]=\left[q\right]$
for any $\lambda\in\mathbb{K},q\in Q$.

The set of all dimensions in $Q$, denoted $Q/{\sim}$, is a finitely
generated free abelian group with multiplication defined by $\left[q\right]\left[q'\right]=\left[qq'\right]$
and identity $\left[1_{Q}\right]$. Hence, $\left[q^{k}\right]=\left[q\right]^{k}$
for any $q\in Q$ and any integer $k$ when $q$ is invertible, or
any non-negative integer otherwise.

The quantities in a dimension $\mathsf{C}\in Q/{\sim}$ form a one-dimensional
vector space over $\mathbb{K}$ with addition inherited from $\mathbb{K}$
\cite[Definition 2.35]{jo3} and a unique zero vector or \emph{zero}
\emph{quantity} $0_{\mathsf{C}}\neq1_{Q}$. While $0_{\mathsf{C}}q=0_{\mathsf{C}\left[q\right]}$
for every $q\in Q$, the product of non-zero quantities is a non-zero
quantity. A quantity is invertible if and only if it is non-zero,
and any non-zero $u\in\mathsf{C}$ is a \emph{unit quantity} for $\mathsf{C}$,
meaning that for every $q\in\mathsf{C}$ there is a unique $\mu\in\mathbb{K}$
for $u$ such that $q=\mu\cdot u$, where $\mu=0$ if and only if
$q=0_{\mathsf{C}}$.

\subsection{Motivation and preview}

Dimensional analysis is based on the principle that when quantities
are represented by scalars relative to a choice of units, so that
a given \emph{relation} between quantities is represented by a particular
relation between scalars relative to a choice of units, that relation
between scalars must hold for any choice of units. For example, if
$q$ and $q'$ are quantities, $\mu_{E}\left(q\right)$ and $\mu_{E}\left(q'\right)$
are scalars representing $q$ and $q'$, respectively, relative to
a basis $E$, and $\mu_{E}\left(q\right)=\mu_{E}\left(q'\right)$
represents the relation $q=q'$ relative to $E$ then $\mu_{E'}\left(q\right)=\mu_{E'}\left(q'\right)$
for any choice of basis $E'$. Below, we define relations between
quantities by equations of the form $q_{0}=\Phi\left(q_{1}.\ldots.q_{n}\right)$,
where $\Phi$ is a quantity function, and relations between scalars
by equations of the form $\mu_{E}\left(q_{0}\right)=\phi\left(\mu_{E}\left(q_{1}\right),\ldots,\mu_{E}\left(q_{n}\right)\right)$,
where $\phi$ is a scalar function. If $\phi$ conforms with the representation
principle just described, it is said to be a ``covariant scalar representation''
of $\Phi$ (Section \ref{s24}).

In Section \ref{sec22}, we consider several closely related notions
shown in the table below:

\smallskip{}
\begin{center}
\begin{tabular}{ccc}
\hline 
 & Basis & Prebasis\tabularnewline
\hline 
Quantity & Basis of quantities & Prebasis of quantities\tabularnewline
Dimension & Basis of dimensions & Prebasis of dimensions\tabularnewline
\hline 
\end{tabular} 
\par\end{center}

\medskip{}
One may question whether these are distinctions that need to be emphasized,
but it turns out that they are quite important. In particular, these
distinctions serve a proof strategy often used in mathematics: results
are proved in one domain in order to be transported to another one,
exploiting the fact that the first domain is better understood or
allows simpler arguments.

In a free abelian group $G$, every maximal independent set of elements
has the same number of elements, the rank $r$ of $G$. In particular,
every basis $\mathsf{E}$ for $G$ has $r$ elements. Thus, every
$\mathsf{D}\in G$ has an expansion $\mathsf{D}=\prod_{i=1}^{r}\mathsf{E}_{i}^{K_{i}}$,
where $\mathsf{E}_{i}\in\mathsf{E}$ for $i=1,\ldots,r$ and $K_{i}$
are unique integers. However, not every maximal independent set $\mathcal{E}=\left\{ \mathcal{E}_{1},\ldots,\mathcal{E}_{r}\right\} $
of $r$ elements of $G$ is a basis; we only have that $\mathsf{D}^{K}=\prod_{i=1}^{r}\mathcal{E}{}_{i}^{K_{i}}$
with unique integers $K,K_{i}$ such that $K>0$ and $\gcd\left(K,K_{1},\ldots,K_{r}\right)=1$
(Section \ref{sec22}). A maximal independent subset of $G$ is a
``prebasis'' for $G$; this is a basis if and only if $K=1$.

In classical dimensional analysis, one considers a free abelian group
$G$ given by a fixed basis such as $\mathsf{E}_{G}=\left\{ \mathsf{L},\mathsf{T},\mathsf{M}\right\} $
(length, time and mass). In the present approach, we start instead
from a free abelian group $G$ without a choice of basis and a quantity
function $\Phi:\mathsf{C}_{1}\times\cdots\times\mathsf{C}_{n}\rightarrow\mathsf{C}_{0}^{\mathfrak{K}}$,
where $\mathsf{C}_{i}\in G$. We then look for those subsets $\mathcal{E}$
of $\left\{ \mathsf{C}_{1},\ldots.\mathsf{C}_{n}\right\} $ which
are prebases for the free abelian group $G'\subseteq G$ generated
by $\left\{ \mathsf{C}_{0},\ldots,\mathsf{C}_{n}\right\} $. At least
one such $\mathcal{E}$ must exist for dimensional analysis to be
possible, but more than one may be found. Each such prebasis may or
may not be a basis for $G'$. 

It is easiest to prove an analogue of the $\pi$ theorem \textendash{}
a representation theorem for $\Phi$ \textendash{} in the case when
$\mathcal{E}$ is a basis for $G'$. In Section \ref{sec:3}, we first
prove a representation theorem that concerns this special case (Theorem
\ref{pimain}), and then generalize to the case when $\mathcal{E}$
is a prebasis (Theorem \ref{pimain-1}) by means of Lemma \ref{t1}.

Informally, a dimension is a quantity that has ``forgotten its size'',
so quantities and dimensions are largely equivalent, but as mathematical
objects dimensions are better understood, being elements of free abelian
groups. Thus, in Section \ref{sec22} some notions are first developed
for dimensions and then carried over to quantities; the main ``transport
theorem'' is Lemma \ref{lem:22}. To prove Theorem \ref{pimain},
we use the covariance principle described above, properties of $\mu_{E}$
and of free abelian groups, and Lemma \ref{lem:22}.

\subsection{\label{sec22}Quantity functions with associated prebases and bases}

A \emph{quantity function} on a quantity space $Q$ is a function
of the form
\begin{equation}
\Phi:\mathsf{C}_{1}\times\cdots\times\mathsf{C}_{n}\rightarrow\mathsf{C}_{0}^{\mathfrak{K}},\qquad\left(q_{1},\ldots,q_{n}\right)\mapsto q_{0}^{\mathfrak{K}}\qquad(n\geq0,\mathsf{C}_{i}\in Q/{\sim}),\label{eq:qfunc}
\end{equation}
where $\mathfrak{K}$ is a positive integer. Equation (\ref{eq:pi1})
now takes the form $q_{0}^{\mathfrak{K}}=\Phi\left(q_{1},\ldots,q_{n}\right)$,
where $q_{0},\ldots,q_{n}$ are quantities. It is easy to show that
$q_{0}^{\mathfrak{K}}\in\mathsf{C}_{0}^{\mathfrak{K}}$ implies $q_{0}\in\mathsf{C}_{0}$
as $Q/{\sim}$ is free abelian.

The dimensions $\mathsf{D}_{1},\ldots,\mathsf{D}_{m}\in Q/{\sim}$
or quantities $q_{1},\ldots,q_{m}\in Q$ are said to be \emph{dependent}
when there are integers $k_{1},\ldots,k_{m}$, not all equal to zero,
such that $\mathsf{D}_{1}^{k_{1}}\cdots\mathsf{D}_{m}^{k_{m}}=\left[1_{Q}\right]$
or $\left[q_{1}^{k_{1}}\cdots q_{m}^{k_{m}}\right]=\left[1_{Q}\right]$,
respectively. Dimensions or quantities which are not dependent are
said to be \emph{independent}. 

A\emph{ prebasis of dimensions} for $\Phi$ is a set $\mathsf{E}=\left\{ \mathsf{E}_{1},\ldots,\mathsf{E}_{r}\right\} \subseteq\left\{ \mathsf{C}_{1},\ldots,\mathsf{C}_{n}\right\} $
such that $\mathsf{E}$ is a maximal independent subset of $\mathsf{F}=\left\{ \mathsf{C}_{0},\mathsf{C}_{1},\ldots,\mathsf{C}_{n}\right\} $.
This means that $\mathsf{E}_{1},\ldots,\mathsf{E}_{r}$ are independent
and for $i=0,\ldots,n$ there are integer sequences $k_{i},k_{i1},\ldots,k_{ir}$
such that 
\begin{equation}
\mathsf{C}_{i}^{k_{i}}=\prod_{j=1}^{r}\nolimits\mathsf{E}_{j}^{k_{ij}}\qquad\left(k_{i}\neq0\right).\label{eq:221}
\end{equation}
A quantity function for which a prebasis of dimensions exists is said
to be \emph{precomplete}. 

Similarly, a \emph{prebasis} (of quantities) for $\Phi$ is a set
of non-zero independent quantities $E=\left\{ e_{1},\ldots,e_{r}\right\} $
such that for $i=0,\ldots,n$ there are integer sequences $k_{i}',k_{i1}',\ldots,k_{ir}'$
such that 
\begin{equation}
\left[q_{i}^{k_{i}'}\right]=\left[\prod_{j=1}^{r}\nolimits e_{j}^{k_{ij}'}\right]\qquad\left(k_{i}'\neq0\right).\label{eq:222}
\end{equation}

It can be shown \cite{jo2} that for each $i$ there is a unique sequence
of \emph{minimal exponent}s $K_{i},K_{i1},\ldots,K_{ir}$ such that
$K_{i}>0$ and $\mathrm{gcd}\left(K_{i},K_{i1},\ldots,K_{ir}\right)=1$
in the set of all sequences of exponents $k_{i},k_{i1},\ldots,k_{ir}$
satisfying (\ref{eq:221}). Similarly, there is a unique sequence
of such minimal exponents $K_{i}',K_{i1}',\ldots,K_{ir}'$ in the
set of all sequences of exponents $k_{i}',k_{i1}',\ldots,k_{ir}'$
satisfying (\ref{eq:222}).

Let $\mathsf{E}=\left\{ \mathsf{E}_{1},\ldots,\mathsf{E}_{r}\right\} $
be a prebasis of dimensions for $\Phi$. If $E=\left\{ e_{1},\ldots,e_{r}\right\} $
and $0_{\mathsf{E}_{j}}\neq e_{j}\in\mathsf{E}_{j}$ for $j=1,\ldots.r$
(henceforth $E\simeq\mathsf{E}$) then (\ref{eq:221}) implies (\ref{eq:222}),
with $K_{i}'=K_{i}$ and $K_{ij}'=K_{ij}$, since $\mathsf{C}_{i}^{K_{i}}=\prod_{j=1}^{r}\nolimits\mathsf{E}_{j}^{K_{ij}}$
implies $\left[q_{i}^{K_{i}}\right]=\left[q_{i}\right]^{K_{i}}=\prod_{j=1}^{r}\left[e_{j}\right]^{K_{ij}}=\left[\prod_{j=1}^{r}e_{j}^{K_{ij}}\right]$.
Also, $e_{1},\ldots,e_{r}$ are independent since $\left[e_{1}\right]\ldots\left[e_{r}\right],$
are independent. Thus, $E$ is a prebasis for $\Phi$.

A prebasis of dimensions for $\Phi$ with associated minimal exponents
such that $K_{i}=1$ for $i=0,\ldots,n$ is said to be a\emph{ basis
of dimensions} for $\Phi$. Similarly, a prebasis (of quantities)
for $\Phi$ such that $K_{i}'=1$ for $i=0,\ldots,n$ is said to be
a\emph{ basis} for $\Phi$. We conclude that if $\mathsf{E}$ is a
 basis of dimensions for $\Phi$ and $E\simeq\mathsf{E}$, then $E$
is a  basis for $\Phi$ with $K_{i}'=K_{i}=1$ and $K_{ij}'=K_{ij}$.
A quantity function with a  basis of dimensions, and thus also a  basis,
is said to be \emph{complete}. 

Recall that all finite maximal independent subsets of a free abelian
group (including all finite bases) have the same number of elements,
the \emph{rank} of the group.
\begin{lem}
\label{lem:L21}Let $\Phi:\mathsf{C}_{1}\times\cdots\times\mathsf{C}_{n}\rightarrow\mathsf{C}_{0}^{\mathfrak{K}}$
be a quantity function on $Q$. Then the subgroup $\left(Q/{\sim}\right)'$
of $Q/{\sim}$ generated by  $\mathsf{F}=\left\{ \mathsf{C}_{0},\ldots,\mathsf{C}_{n}\right\} $
is free abelian, and the number of elements in any prebasis of dimensions
for $\Phi$ is equal to the rank of $\left(Q/{\sim}\right)'$.
\end{lem}

\begin{proof}[Proof (sketch)]
Any maximal independent subset of $\mathsf{F}$ is a finite maximal
independent subset of $\left(Q/{\sim}\right)'$. This proves the assertion
since $\left(Q/{\sim}\right)'$, a subgroup of $Q/{\sim}$, is also
free abelian.
\end{proof}
\begin{lem}
\label{lem:22}Let $\mathsf{E}$ be a basis of dimensions for $\Phi:\mathsf{C}_{1}\times\cdots\times\mathsf{C}_{n}\rightarrow\mathsf{C}_{0}^{\mathfrak{K}}$,
$E=\left\{ e_{1},\ldots,e_{r}\right\} \simeq\mathsf{E}$ a basis for
$\Phi$, and $q_{i}\in\mathsf{C}_{i}$. If $\left[q_{i}\right]=\prod_{j=1}^{r}\nolimits\left[e_{i}\right]^{K_{ij}}$
then $q_{i}$ has a unique expansion $q_{i}=\mu\cdot\prod_{j=1}^{r}e_{j}^{K_{ij}}$.
\end{lem}

\begin{proof}
Since $q_{i}\in\left[\prod_{j=1}^{r}e_{j}^{K_{ij}}\right]$ and $\prod_{j=1}^{r}\nolimits e_{j}^{K_{ij}}\neq0_{\left[q_{i}\right]}$,
there is a unique $\mu\in\mathbb{K}$ for $\prod_{j=1}^{r}e_{j}^{K_{ij}}$
such that $q_{i}=\mu\cdot\prod_{j=1}^{r}e_{j}^{K_{ij}}$. Also, if
$\mu'\cdot\prod_{j=1}^{r}e_{j}^{k_{ij}}=\mu\cdot\prod_{j=1}^{r}e_{j}^{K_{ij}}$
then $\prod_{j=1}^{r}\left[e_{j}\right]^{k_{ij}}=\prod_{j=1}^{r}\left[e_{j}\right]^{K_{ij}}$,
so $\prod_{j=1}^{r}\left[e_{j}\right]^{k_{ij}-K_{ij}}=\left[1_{Q}\right],$
so $k_{ij}=K_{ij}$ for $j=1,\ldots,r$ since $\left[e_{1}\right],\ldots,\left[e_{r}\right]$
are independent, so $\mu'=\mu$.
\end{proof}
Let $Q'$ be the subspace of $Q$ generated by all non-zero $q_{0},\ldots,q_{n}$,
where $q_{i}\in\mathsf{C}_{i}$. Since $e_{1},\ldots,e_{r}$ are independent
and generate $Q'$ by Lemma \ref{lem:22}, every $q\in Q'$ has a
unique expansion of the form $q=\mu_{E}\left(q\right)\cdot\prod_{j=1}^{r}e_{j}^{K_{j}}$
in terms of the ``local basis'' $E$. For any $E$, $\mu_{E}\left(qq'\right)=\mu_{E}\left(q\right)\mu_{E}\left(q'\right)$
for any $q,q'\in Q'$, and $\mu_{E}\left(q^{-1}\right)=\mu_{E}\left(q\right)^{-1}$
for any invertible $q\in Q'$ \cite{jo3}. If $q\in\left[1_{Q}\right]$
then $q=\mu_{E}\left(q\right)\cdot\prod_{i=1}^{n}e_{i}^{0}$, so in
that case $\mu_{E}\left(q\right)$ does not depend on $E$. 

\subsection{\label{s24}Covariant scalar representations}

The ``physically meaningful'' quantity functions of interest in
augmented dimensional analysis have scalar representations that do
not depend on a choice of basis. Specifically, a \emph{covariant scalar
representation} of a complete quantity function $\Phi$ of the form
(\ref{eq:qfunc}) is a function $\phi:\mathbb{K}^{n}\rightarrow\mathbb{K}$
such that 
\begin{equation}
\mu_{E}\left(\Phi\left(q_{1},\ldots,q_{n}\right)\right)=\phi\left(\mu_{E}\left(q_{1}\right),\ldots,\mu_{E}\left(q_{n}\right)\right)\label{eq:covscalfunc-1}
\end{equation}
for any basis $E$ for $\Phi$ and any $q_{1},\ldots,q_{n}$.

If $E$ is only allowed to be any basis such that $E\simeq\mathsf{E}$,
where $\mathsf{E}$ is a fixed  basis of dimensions, then $\phi$
is said to be a\emph{ scaling-covariant} scalar representation. A
covariant scalar representation is obviously scaling-covariant.

Note that the identity map $\mathrm{id}:\mathbb{K}\rightarrow\mathbb{K}$
is a covariant scalar representation of the identity map $\mathrm{Id}:\mathsf{C}\rightarrow\mathsf{C}$
since $\mu_{E}\left(\mathrm{Id}\left(q\right)\right)=\mathrm{id}\left(\mu_{E}\left(q\right)\right)$
for all $q$ and $E$. Also, if $\phi_{1}$ is a covariant scalar
representation of $\Phi_{1}:\mathsf{C}_{1}\rightarrow\mathsf{D}_{1}$,
and $\phi_{2}$ of $\Phi_{2}:\mathsf{C_{2}}\rightarrow\mathsf{D}_{2}$,
then $\phi_{1}\phi_{2}$ is a covariant scalar representation of $\Phi_{1}\Phi_{2}$
since 
\[
\mu_{E}\left(\Phi_{1}\left(q_{1}\right)\Phi_{2}\left(q_{2}\right)\right)=\mu_{E}\left(\Phi_{1}\left(q_{1}\right)\right)\mu_{E}\left(\Phi_{2}\left(q_{2}\right)\right)=\phi_{1}\left(\mu_{E}\left(q_{1}\right)\right)\phi_{2}\left(\mu_{E}\left(q_{2}\right)\right)
\]
for all $q_{1},q_{2}$ and $E$, and if $\phi$ is a covariant scalar
representation of $\Phi:\mathsf{D}_{1}\times\cdots\times\mathsf{D}_{n}\rightarrow\mathsf{D}_{0}$
and $\omega_{1},\ldots,\omega_{n}$ are covariant scalar representations
of $\text{\ensuremath{\Omega}}_{1}:\mathsf{C}_{1}\rightarrow\mathsf{D}_{1},\ldots,\text{\ensuremath{\Omega}}_{n}:\mathsf{C}_{n}\rightarrow\mathsf{D}_{n}$,
respectively, then $\phi\circ\left(\omega_{1},\ldots,\omega_{n}\right)$
is a scalar representation of $\Phi\circ\left(\Omega_{1},\ldots,\Omega_{n}\right)$
since
\[
\mu_{E}\left(\Phi\left(\Omega_{1}\left(q_{1}\right),\ldots,\Omega_{n}\left(q_{n}\right)\right)\right)=\phi\left(\mu_{E}\left(\Omega_{1}\left(q_{1}\right)\right),\ldots,\mu_{E}\left(\Omega_{n}\left(q_{n}\right)\right)\right)=\phi\left(\omega_{1}\left(\mu_{E}\left(q_{1}\right)\right),\ldots,\omega_{n}\left(\mu_{E}\left(q_{n}\right)\right)\right)
\]
 for all $q_{1},\ldots,q_{n}$ and $E$. These results apply to scaling-covariant
scalar representations as well.

\textcolor{black}{Note that a complete quantity function need not
have a covariant scalar representation. }
\begin{example}
\textcolor{black}{\label{ex1}Consider a quantity function
\[
\Phi_{u}:\mathsf{C}\rightarrow\left[1_{Q}\right],\qquad\Phi_{u}\left(\lambda\cdot u\right)\mapsto\lambda\cdot1_{Q}\qquad\left(0_{\mathsf{C}}\neq u\in\mathsf{C},\mathsf{C}\neq\left[1_{Q}\right]\right).
\]
$\Phi_{u}$ has a  basis of dimensions $\left\{ \mathsf{C}\right\} $
since $\mathsf{C}=\mathsf{C}^{1}$ and $\left[1_{Q}\right]=\mathsf{\mathsf{C}^{0}}$.
Let us show that
\[
\mu_{\left\{ u\right\} }\left(\Phi_{u}\left(u\right)\right)\neq\phi\left(\mu_{\left\{ u\right\} }\left(u\right)\right)\quad\mathrm{or}\quad\mu_{\left\{ 2\cdot u\right\} }\left(\Phi_{u}\left(2\cdot u\right)\right)\neq\phi\left(\mu_{\left\{ 2\cdot u\right\} }\left(2\cdot u\right)\right).
\]
We have $\mu_{\left\{ u\right\} }\left(\Phi_{u}\left(u\right)\right)=1$
since $\Phi_{u}\left(u\right)=1\cdot1_{Q}$, but $\mu_{\left\{ 2\cdot u\right\} }\left(\Phi_{u}\left(2\cdot u\right)\right)=2$
since $\mu_{\left\{ u\right\} }\left(\Phi_{u}\left(2\cdot u\right)\right)=2$
and $\Phi_{u}\left(x\right)\in\left[1_{Q}\right]$. However, $\phi\left(\mu_{\left\{ u\right\} }\left(u\right)\right)=\phi\left(\mu_{\left\{ 2\cdot u\right\} }\left(2\cdot u\right)\right)=\phi\left(1\right)$,
leading to the desired result. }

\textcolor{black}{Conversely, if $\Phi:\mathsf{C}\rightarrow\left[1_{Q}\right]$
has a scaling-covariant scalar representation then $\Phi$ is a constant
function, as shown in Example }\ref{x3-1}\textcolor{black}{.}
\end{example}

\section{\label{sec:3}Representation theorems}

Given a quantity function $\Phi:\mathsf{C}_{1}\times\cdots\times\mathsf{C}_{n}\rightarrow\mathsf{C}_{0}^{\mathfrak{K}}$
and a permutation $\sigma$ of $\left\{ 1,\ldots,n\right\} $ one
can construct a new quantity function $\Phi':\mathsf{C}_{\sigma\left(1\right)}\times\cdots\times\mathsf{C}_{\sigma\left(n\right)}\rightarrow\mathsf{C}_{0}^{\mathfrak{K}}$
by setting $\Phi'\left(q_{\sigma\left(1\right)},\ldots,q_{\sigma\left(n\right)}\right)=\Phi\left(q_{1},\ldots,q_{n}\right)$.
One may in particular reorder the arguments so that the elements of
a certain prebasis for $\Phi$ come first in $\left(q_{\sigma\left(1\right)},\ldots,q_{\sigma\left(n\right)}\right)$;
we call this a\emph{ prebasis reordering} of $\left(q_{1},\ldots,q_{n}\right)$.
The results in this section apply to a quantity function obtained,
if necessary, by a prebasis reordering of arguments.
\begin{thm}[special representation theorem]
\label{pimain}Let $Q$ be a quantity space, let 
\[
\Phi:\mathsf{E}_{1}\times\cdots\times\mathsf{E}_{r}\times\mathsf{D}_{1}\times\cdots\times\mathsf{D}_{n-r}\rightarrow\mathsf{D}_{0},\qquad\left(x_{1},\ldots,x_{r}\right)\left(y_{1},\ldots,y_{n-r}\right)\mapsto y_{0}\qquad\left(r,n-r\geq0\right)
\]
be a complete quantity function on $Q$ and let $\mathsf{E}=\left\{ \mathsf{E}_{1},\ldots,\mathsf{E}_{r}\right\} $
be a basis of dimensions for $\Phi$ with associated minimal exponents
$1,K_{ij}$ such that $\mathsf{D}_{i}=\prod_{j=1}^{r}\mathsf{E}_{j}^{K_{ij}}$
for $i=0,\ldots,n-r$. If $\Phi$ has a scaling-covariant scalar representation,
then there exists a quantity function of $n-r$ arguments
\[
\Psi:\left[1_{Q}\right]\times\cdots\times\left[1_{Q}\right]\rightarrow\left[1_{Q}\right]
\]
such that if $x_{1},\ldots,x_{r}$ are non-zero then
\begin{equation}
\pi_{0}=\Psi\left(\pi_{1},\ldots,\pi_{n-r}\right),\label{eq:pitheorem-1}
\end{equation}
where $\pi_{i}=y_{i}\prod_{j=1}^{r}\nolimits x_{j}^{-K_{ij}}$ for
$i=0,\ldots,n-r$, or equivalently
\begin{equation}
\Phi\left(x_{1},\ldots,x_{r}\right)\left(y_{1},\ldots,y_{n-r}\right)=y_{0}=\prod_{j=1}^{r}\nolimits\!x_{j}^{K_{0j}}\,\Psi\left(\pi_{1},\ldots,\pi_{n-r}\right).\label{eq:pitheorem2-1}
\end{equation}
\end{thm}

\begin{proof}
Fix $X=\left\{ x_{1},\ldots,x_{r}\right\} ${\small{} such that $X\simeq\mathsf{E}$.
}Any $E=\left\{ e_{1},\ldots,e_{r}\right\} $ such that $E\simeq\mathsf{E}$
is a basis for $\Phi$, and $x_{i}\in\left[x_{i}\right]=\left[e_{i}\right]$
so by Lemma \ref{lem:22} there are unique expansions relative to
$E$,
\begin{equation}
x_{i}=\mu_{E}\left(x_{i}\right)\cdot e_{i}\qquad\left(i=1,\ldots,r\right),\qquad\qquad y_{i}=\mu_{E}\left(y_{i}\right)\cdot\prod_{j=1}^{r}\nolimits e_{j}^{K_{ij}}\qquad\left(i=0,\ldots,n-r\right).\label{eq:yi-1-1}
\end{equation}
{\small{}Set
\begin{equation}
\breve{y}{}_{i}=\prod_{j=1}^{r}\nolimits x_{j}^{K_{ij}}\qquad\left(i=0,\ldots,n-r\right)\label{eq:yi-2-1}
\end{equation}
so that $\breve{y}{}_{i}\neq0_{\left[\breve{y}{}_{i}\right]}$ since}
$x_{j}\neq0_{\left[x_{j}\right]}$ for $j=1,\ldots r${\small{}. Then
$\pi_{i}=y_{i}\breve{y}_{i}^{-1}$ by definition, and (}\ref{eq:yi-1-1})
and (\ref{eq:yi-2-1}{\small{}) gives
\[
\left[y_{i}\right]=\left[\mu_{E}\left(y_{i}\right)\cdot\prod_{j=1}^{r}\nolimits e_{j}^{K_{ij}}\right]=\left[\prod_{j=1}^{r}\nolimits e_{j}^{K_{ij}}\right]=\left[\prod_{j=1}^{r}\nolimits x_{j}^{K_{ij}}\right]=\left[\breve{y}{}_{i}\right]
\]
since $\left[x_{j}\right]=\left[e_{j}\right]$}, so {\small{}$\pi_{i}\in\left[y_{i}\breve{y}_{i}^{-1}\right]=\left[y_{i}\right]\left[\breve{y}_{i}\right]^{-1}=\left[1_{Q}\right]$},
so $\mu_{E}\left(\pi_{i}\right)$ does not depend on $E$.

Let $\mathbf{q}$ denote the sequence of quantities $\left(x_{1},\ldots,x_{r}\right)\left(y_{1},\ldots,y_{n-r}\right)$,
and let $\mu_{E}\left(\mathbf{q}\right)$ be the sequence of scalars
$\left(\mu_{E}\left(x_{1}\right),\ldots,\mu_{E}\left(x_{r}\right)\right)\left(\mu_{E}\left(y_{1}\right),\ldots,\mu_{E}\left(y_{n-r}\right)\right)$.
By definition, $y_{0}=\Phi\left(\mathbf{q}\right)$ and by assumption
there is a function $\phi:\mathbb{K}^{n}\rightarrow\mathbb{K}$ such
that $\mu_{E}\left(\Phi\left(\mathbf{q}\right)\right)=\phi\left(\mu_{E}\left(\mathbf{q}\right)\right)$
for any $\mathbf{q}$ and $E\simeq\mathsf{E}$. Also, $\mu_{E}\left(\breve{y}_{0}\right)\neq0$
since $\breve{y}{}_{0}\neq0_{\left[\breve{y}{}_{0}\right]}$, and
$\mu_{E}\left(\prod_{j=1}^{r}x_{j}^{K_{0j}}\right)=\prod_{j=1}^{r}\mu_{E}\left(x_{j}\right)^{K_{0j}}$.
There is thus a function $\varphi:\mathbb{K}^{n}\rightarrow\mathbb{K}$
such that for any $\mathbf{q}$ and $E\simeq\mathsf{E}$ we have 
\begin{gather*}
\mu_{E}\left(\pi_{0}\right)=\mu_{E}\left(y_{0}\breve{y}_{0}^{-1}\right)=\frac{\mu_{E}\left(\Phi\left(\mathbf{q}\right)\right)}{\mu_{E}\left(\breve{y}_{0}\right)}=\frac{\phi\left(\mu_{E}\left(\mathbf{q}\right)\right)}{\prod_{j=1}^{r}\mu_{E}\left(x_{j}\right)^{K_{0j}}}=\varphi\left(\mu_{E}\left(\mathbf{q}\right)\right).
\end{gather*}

Furthermore, as $\mu_{E}\left(\pi_{i}\right)=\mu_{E}\left(y_{i}\right)/\prod_{j=1}^{r}\mu_{E}\left(x_{j}\right)^{K_{ij}}$
for $i=1,\ldots,n-r$ there is a function
\begin{gather*}
\omega:\mu_{E}\left(\mathbf{q}\right)\:\mapsto\:\left(\mu_{E}\left(x_{1}\right),\ldots,\mu_{E}\left(x_{r}\right)\right)\left(\mu_{E}\left(\pi_{1}\right),\ldots,\mu_{E}\left(\pi_{n-r}\right)\right),
\end{gather*}
and $\omega$ is bijective since conversely $\mu_{E}\left(y_{i}\right)=\mu_{E}\left(\pi_{i}\right)\prod_{j=1}^{r}\mu_{E}\left(x_{j}\right)^{K_{ij}}$
for $i=1,\ldots,n-r$. Hence, there is a function $\Gamma=\varphi\circ\omega^{-1}:\mathbb{K}^{n}\rightarrow\mathbb{K}$
such that
\[
\mu_{E}\left(\pi_{0}\right)=\varphi\left(\mu_{E}\left(\mathbf{q}\right)\right)=\Gamma\left(\mu_{E}\left(x_{1}\right),\ldots,\mu_{E}\left(x_{r}\right)\right)\left(\mu_{E}\left(\pi_{1}\right),\ldots,\mu_{E}\left(\pi_{n-r}\right)\right).
\]

Note that we can set $E=X$ since $X\simeq\mathsf{E}$. There is thus
a function $\psi:\mathbb{K}^{n-r}\rightarrow\mathbb{K}$ such that
\begin{gather}
\mu_{X}\!\left(\pi_{0}\right)=\Gamma\left(1,\ldots,1\right)\left(\mu_{X}\!\left(\pi_{1}\right),\ldots,\mu_{X}\!\left(\pi_{n-r}\right)\right)=\psi\left(\mu_{X}\!\left(\pi_{1}\right),\ldots,\mu_{X}\!\left(\pi_{n-r}\right)\right)\label{eq:scalpi1}
\end{gather}
since $x_{j}=1\cdot x_{j}$ so that $\mu_{X}\left(x_{j}\right)=1$
for $j=1,\ldots,r$.

To complete the proof, we convert the scalar function $\psi$, obtained
from the quantity function $\Phi$, back into a quantity function
$\Psi$. As $\mu_{E}\left(\pi_{i}\right)$ does not depend on $E\simeq\mathsf{E}$,
we can define a quantity function of $n-r$ arguments 
\[
\Psi:\left[1_{Q}\right]\times\cdots\times\left[1_{Q}\right]\rightarrow\left[1_{Q}\right],
\]
which depends only on $\psi$, by setting
\[
\Psi\left(\mu_{X}\!\left(\pi_{1}\right)\cdot1_{Q},\ldots,\mu_{X}\!\left(\pi_{n-r}\right)\cdot1_{Q}\right)=\psi\left(\mu_{X}\!\left(\pi_{1}\right),\ldots,\mu_{X}\!\left(\pi_{n-r}\right)\right)\cdot1_{Q},
\]
so that, by (\ref{eq:scalpi1}),
\begin{equation}
\mu_{X}\left(\pi_{0}\right)\cdot1_{Q}=\Psi\left(\mu_{X}\left(\pi_{1}\right)\cdot1_{Q},\ldots,\mu_{X}\left(\pi_{n-r}\right)\cdot1_{Q}\right).\label{eq:prepieq}
\end{equation}
Recall that $\pi_{i}=\mu_{X}\left(\pi_{i}\right)\cdot1_{Q}$ is the
unique expansion of $\pi_{i}$ relative to $X$ since $\pi_{i}\in\left[1_{Q}\right]$,
so we can rewrite (\ref{eq:prepieq}) as 
\[
\pi_{0}=\Psi\left(\pi_{1},\ldots,\pi_{n-r}\right).
\]
We have thus derived (\ref{eq:pitheorem-1}), or equivalently (\ref{eq:pitheorem2-1}),
proving the theorem.
\end{proof}
As {\small{}$\breve{y}{}_{0}\neq0_{\left[\breve{y}{}_{0}\right]}$,}
$\text{\ensuremath{\Phi\left(x_{1},\ldots,x_{r}\right)\left(y_{1},\ldots,y_{n-r}\right)}}=\breve{y}_{0}\Psi\left(\pi{}_{1},\ldots,\pi{}_{n-r}\right)=\breve{y}_{0}\Psi'\left(\pi{}_{1},\ldots,\pi{}_{n-r}\right)$
implies $\Psi=\Psi'$, and {\small{}in view of Theorem }\ref{pimain}
$\Phi\left(q_{1},\ldots,q_{n}\right)=0_{\left[y_{0}\right]}$ if and
only if $\Psi\left(\pi{}_{1},\ldots,\pi{}_{n-r}\right)=0_{\left[1_{Q}\right]}$,
as in Buckingham's original $\pi$ theorem \cite{bu}.
\begin{lem}
\label{t1}\textup{ }\textup{\emph{Let $Q$ be a quantity space, let
\[
\Phi:\mathsf{E}_{1}\times\cdots\times\mathsf{E}_{r}\times\mathsf{D}_{1}\times\cdots\times\mathsf{D}_{n-r}\rightarrow\mathsf{D_{0}^{\mathfrak{K}}},\qquad\left(x_{1},\ldots,x_{r}\right)\left(y_{1},\ldots,y_{n-r}\right)\mapsto y_{0}^{\mathfrak{K}}\qquad\left(r,n-r\geq0\right),
\]
where $y_{0}\in\mathsf{D}_{0}$ and $\mathfrak{K}>0$, be a precomplete
quantity function on $Q$ and let $\mathsf{E}=\left\{ \mathsf{E}_{1},\ldots,\mathsf{E}_{r}\right\} $
be a prebasis of dimensions for $\Phi$ with associated minimal exponents
$K_{i},K_{ij}$ }}such that $\mathsf{D}_{i}^{K_{i}}=\prod_{j=1}^{r}\mathsf{E}_{j}^{K_{ij}}$\textup{\emph{
for $i=0,\ldots,n-r$. If $\mathfrak{K}=K_{0}$ and there exists a
bijection of arguments
\[
\chi:\left(x_{1},\ldots,x_{r}\right)\left(y_{1},\ldots,y_{n-r}\right)\mapsto\left(x_{1},\ldots,x_{r}\right)\left(y_{1}^{K_{1}},\ldots,y_{n-r}^{K_{n-r}}\right)\qquad\left(x_{i}\in\mathsf{E}_{i},y_{i}\in\mathsf{D}_{i}\right),
\]
then there exists a complete quantity function
\[
\Phi^{*}:\mathsf{E}_{1}\times\cdots\times\mathsf{E}_{r}\times\mathsf{D}_{1}^{K_{1}}\times\cdots\times\mathsf{D}_{n-r}^{K_{n-r}}\rightarrow\mathsf{D}_{0}^{\mathfrak{K}},\qquad\left(x_{1},\ldots,x_{r}\right)\left(y_{1}^{K_{1}},\ldots,y_{n-r}^{K_{n-r}}\right)\mapsto y_{0}^{\mathfrak{K}},
\]
with $\left\{ \mathsf{E}_{1},\ldots,\mathsf{E}_{r}\right\} $ a  basis
of dimensions and such that
\begin{equation}
\Phi^{*}\left(x_{1},\ldots,x_{r}\right)\left(y_{1}^{K_{1}},\ldots,y_{n-r}^{K_{n-r}}\right)=\Phi\left(x_{1},\ldots,x_{r}\right)\left(y_{1},\ldots,y_{n-r}\right).\label{eq:11}
\end{equation}
}}
\end{lem}

\begin{proof}
Note that if $y_{i}\in\mathsf{D}_{i}$ then $\left[y_{i}\right]=\mathsf{D}_{i}$,
so $y_{i}^{K_{i}}\in\left[y_{i}^{K_{i}}\right]=\left[y_{i}\right]^{K_{i}}=\mathsf{D}_{i}^{K_{i}}$.
If we set $\Phi^{*}=\Phi\circ\chi^{-1}$ then (\ref{eq:11}) holds,
and $\left\{ \mathsf{E}_{1},\ldots,\mathsf{E}_{r}\right\} $ is a
 basis of dimensions for $\Phi^{*}$ since $\mathsf{E}_{i}=\prod_{j=1}^{r}\nolimits\!\mathsf{E}_{j}^{\delta_{ij}}$
for $i=1,\ldots,r$, \emph{$\mathsf{D}_{i}^{K_{i}}=\prod_{j=1}^{r}\mathsf{E}_{j}^{K_{ij}}$
}for\emph{ $i=1,\ldots,n-r$}, and $\mathsf{D}_{0}^{\mathfrak{K}}=\prod_{j=1}^{r}\mathsf{E}_{j}^{K_{0j}}$.
\end{proof}
If $\mathfrak{K}=K_{0}$ we say that the choice of $\mathfrak{K}$,
or the corresponding equation, is \emph{consistent}, and a precomplete
quantity function $\Phi$ with a bijection $\chi$ as described above
is said to be \emph{portable}. By first applying Lemma \ref{t1} to
$\Phi$ and then applying Theorem \ref{pimain} to $\Phi^{*}$, we
prove a general representation theorem\textcolor{brown}{.}
\begin{thm}[general representation theorem]
\label{pimain-1}\textup{\emph{Let $Q$ be a quantity space, let
\[
\Phi:\mathsf{E}_{1}\times\cdots\times\mathsf{E}_{r}\times\mathsf{D}_{1}\times\cdots\times\mathsf{D}_{n-r}\rightarrow\mathsf{D_{0}^{\mathfrak{K}}},\qquad\left(x_{1},\ldots,x_{r}\right)\left(y_{1},\ldots,y_{n-r}\right)\mapsto y_{0}^{\mathfrak{K}}\qquad\left(r,n-r\geq0\right),
\]
where $y_{0}\in\mathsf{D}_{0}$ and $\mathfrak{K}>0$, be a precomplete
quantity function on $Q$ and let $\mathsf{E}=\left\{ \mathsf{E}_{1},\ldots,\mathsf{E}_{r}\right\} $
be a prebasis of dimensions for $\Phi$ with associated minimal exponents
$K_{i},K_{ij}$ }}such that $\mathsf{D}_{i}^{K_{i}}=\prod_{j=1}^{r}\mathsf{E}_{j}^{K_{ij}}$\textup{\emph{
for $i=0,\ldots,n-r$. If $\Phi$ is consistent and portable then
there exists a complete quantity function
\[
\Phi^{*}:\mathsf{E}_{1}\times\cdots\times\mathsf{E}_{r}\times\mathsf{D}_{1}^{K_{1}}\times\cdots\times\mathsf{D}_{n-r}^{K_{n-r}}\rightarrow\mathsf{D}_{0}^{\mathfrak{K}},\qquad\left(x_{1},\ldots,x_{r}\right)\left(y_{1}^{K_{1}},\ldots,y_{n-r}^{K_{n-r}}\right)\mapsto y_{0}^{\mathfrak{K}},
\]
with $\left\{ \mathsf{E}_{1},\ldots,\mathsf{E}_{r}\right\} $ a basis
of dimensions and such that
\begin{equation}
\Phi^{*}\left(x_{1},\ldots,x_{r}\right)\left(y_{1}^{K_{1}},\ldots,y_{n-r}^{K_{n-r}}\right)=\Phi\left(x_{1},\ldots,x_{r}\right)\left(y_{1},\ldots,y_{n-r}\right),\label{eq:11-1}
\end{equation}
and if $\Phi^{*}$ }}has a scaling-covariant scalar representation,
then there exists a quantity function of $n-r$ arguments
\[
\Psi:\left[1_{Q}\right]\times\cdots\times\left[1_{Q}\right]\rightarrow\left[1_{Q}\right]
\]
such that if $x_{1},\ldots,x_{r}$ are non-zero then
\begin{gather}
\pi_{0}=\Psi\left(\pi_{1},\ldots,\pi_{n-r}\right),\label{eq:piteorem-1-1}
\end{gather}
where $\pi_{i}=y_{i}^{K_{i}}\prod_{j=1}^{r}\nolimits x_{j}^{-K_{ij}}$
for $i=0,\ldots,n-r$, or equivalently
\begin{equation}
\Phi\left(x_{1},\ldots,x_{r}\right)\left(y_{1},\ldots,y_{n-r}\right)=y_{0}^{\mathfrak{K}}=\prod_{j=1}^{r}\nolimits\!x_{j}^{K_{0j}}\,\Psi\left(\pi_{1},\ldots,\pi_{n-r}\right).\label{eq:eq:pitheorem-2}
\end{equation}
\end{thm}

Each of the following conditions guarantees the existence of a bijection
$\chi$ in Lemma \ref{t1}:
\begin{enumerate}
\item $n=r$.
\item $K_{i}=1$ for $i=1,\ldots,n-r$.
\item The restriction of $\chi$ to $\mathsf{D}_{i}$ is bijective for $i=1,\ldots,n-r$;
this generalizes (2).
\end{enumerate}
In proofs of scalar versions of the $\pi$ theorem it is generally
assumed that $\mu_{E}\left(q_{i}\right)\geq0$ for all $q_{i}$. If
$\mathbb{K}=\mathbb{C}$, we need not assume this to prevent roots
of negative numbers from appearing in equations, but another reason
for assuming that $\mu_{E}\left(y_{i}\right)\geq0$ is that then we
can let $\chi^{-1}\left(\mu_{E}\left(y_{i}\right)^{n}\right)$ be
the unique non-negative $n$th root of $\mu_{E}\left(q_{i}\right)^{n}$.
In a version of the $\pi$ theorem for an ordered quantity space $\mathfrak{Q}$,
where the order is given by a choice of basis for $\mathfrak{Q}$
\cite{ra2}, one could similarly assume that $y_{i}\geq0_{\left[y_{i}\right]}$
for all $y_{i}$. However, $\left(x\right)\left(y\right)\mapsto\left(x\right)\left(y^{2}\right)$,
for example, is a bijection also if we require that $y\leq0_{\left[y\right]}$,
so a universal positive sign requirement is too strict and not quite
to the point.

\section{Doing dimensional analysis}

Let $\Phi$ be a quantity function on $Q$, and consider the corresponding
equation 
\begin{equation}
q_{0}^{\mathfrak{K}}=\Phi\left(q_{1},\ldots,q_{n}\right)\qquad\left(n\geq0\right),\label{eq:41}
\end{equation}
where $\mathfrak{K}$ is an initially unspecified positive integer,
Recall that the purpose of augmented dimensional analysis is to represent
(\ref{eq:41}) by a definite system of equations of the form 
\[
q_{0}^{\mathfrak{K}}=\prod_{j=1}^{r}\nolimits x_{j}^{K_{j}}\,\Psi\left(\pi_{1},\ldots,\pi_{n-r}\right),
\]
where $\pi_{i}=y_{i}^{K_{i}}\prod_{j=1}^{r}x_{j}^{-K_{ij}}$ and $r,n-r\geq0$.
Based on the results in Section \ref{sec:3}, we describe in some
detail below when and how this can be done.

\subsection{Systems of quantity equations}

If $\Phi$ is precomplete then we may use one or more  prebasis reorderings
$\left(x_{\theta,1},\ldots,x_{\theta,r}\right)\left(y_{\theta,1},\ldots.y_{\theta,n-r}\right)$
of $\left(q_{1},\ldots,q_{n}\right)$ such that $\left\{ \left[x_{\theta,1}\right],\ldots,\left[x_{\theta,r}\right]\right\} $
is a prebasis of dimensions for $\Phi$ and $E_{\theta}=\left\{ x_{\theta,1},\ldots,x_{\theta,r}\right\} $
a prebasis for $\Phi$ whenever all $x_{\theta,j}$ are non-zero quantities.
Thus, we obtain $\Theta$ equations
\begin{equation}
\begin{cases}
y_{0}^{\mathfrak{K_{\theta}}}=\Phi_{\theta}'\left(x_{\theta,1},\ldots,x_{\theta,r}\right)\left(y_{\theta,1},\ldots,y_{\theta,n-r}\right) & \left(\theta=1,\ldots,\Theta\right),\end{cases}\label{eq:42}
\end{equation}
where $y_{0}=q_{0}$ and $\Phi_{\theta}'\left(x_{\theta,1},\ldots,x_{\theta,r}\right)\left(y_{\theta,1},\ldots,y_{\theta,n-r}\right)=\Phi\left(q_{1},\ldots,q_{n}\right)$
for all $\theta$. For each $E_{\theta}$ we have associated minimal
exponents $K_{\theta,i},K_{\theta,ij}$ such that
\begin{equation}
\left[y_{\theta,i}\right]^{K_{\theta,i}}=\prod_{j=1}^{r}\nolimits\left[x_{\theta,j}\right]^{K_{\theta,ij}}\quad\left(i=0,\ldots,n-r\right).\label{eq:421}
\end{equation}

We obtain $\Theta$ consistent equations of the form (\ref{eq:42})
by setting $\mathfrak{K}_{\theta}=K_{\theta,0}$ for all $\theta$.
If each $\Phi_{\theta}'$ is portable as well, meaning that there
are bijections
\[
\left(x_{\theta,1},\ldots,x_{\theta,r}\right)\left(y_{\theta,1},\ldots,y_{\theta,n-r}\right)\mapsto\left(x_{\theta,1},\ldots,x_{\theta,r}\right)\left(y_{\theta,1}^{K_{\theta,1}},\ldots,y_{\theta,n-r}^{K_{\theta,n-r}}\right)\quad\left(\theta=1,\ldots,\Theta\right),
\]
 then there are $\Theta$ equations with complete functions $\Phi_{\theta}^{*}$,
\begin{equation}
\begin{cases}
y_{0}^{\mathfrak{K_{\theta}}}=\Phi_{\theta}^{*}\left(x_{\theta,1},\ldots,x_{\theta,r}\right)\left(y_{\theta,1}^{K_{\theta,1}},\ldots,y_{\theta,n-r}^{K_{\theta,n-r}}\right) & \left(\theta=1,\ldots,\Theta\right),\end{cases}\label{eq:43}
\end{equation}
 where $\Phi_{\theta}^{*}\left(x_{\theta,1},\ldots,x_{\theta,r}\right)\left(y_{\theta,1}^{K_{\theta,1}},\ldots,y_{\theta,n-r}^{K_{\theta,n-r}}\right)=\Phi_{\theta}'\left(x_{\theta,1},\ldots,x_{\theta,r}\right)\left(y_{\theta,1},\ldots,y_{\theta,n-r}\right)$
for all $\theta$.

If each $\Phi_{\theta}^{*}$ has a scaling-covariant scalar representation
then we obtain $\Theta$ equations
\begin{equation}
\begin{cases}
y_{0}^{\mathfrak{K}_{\theta}}=\prod_{j=1}^{r}x_{\theta,j}^{K_{\theta,0j}}\,\Psi_{\theta}\left(\pi_{\theta,1},\ldots,\pi_{\theta,n-r}\right) & \left(\theta=1,\ldots,\Theta\right),\end{cases}\label{eq:set2}
\end{equation}
 where $\prod_{j=1}^{r}x_{\theta,j}^{K_{\theta,0j}}\,\Psi_{\theta}\left(\pi_{\theta,1},\ldots,\pi_{\theta,n-r}\right)=\Phi_{\theta}^{*}\left(x_{\theta,1},\ldots,x_{\theta,r}\right)\left(y_{\theta,1}^{K_{\theta,1}},\ldots,y_{\theta,n-r}^{K_{\theta,n-r}}\right)$
for all $\theta$. 

Finally, since $\Psi_{\theta}\left(\pi_{\theta,1},\ldots,\pi_{\theta,n-r}\right)\in\left[1_{Q}\right]$
for all $\theta$ we obtain from (\ref{eq:set2}) $\Theta$ equations
\begin{equation}
\begin{cases}
y_{0}^{\mathfrak{K}}=y_{0}^{C_{\theta}\mathfrak{K}_{\theta}}=\prod_{j=1}^{r}\nolimits x_{\theta,j}^{C_{\theta}K_{\theta,0j}}\,\Psi_{\theta}^{C_{\theta}}\left(\pi_{\theta,1},\ldots,\pi_{\theta,n-r}\right) & \left(\theta=1,\ldots,\Theta\right),\end{cases}\label{eq:set2b}
\end{equation}
where $C_{\theta}=\mathrm{lcm}\left(\mathfrak{K}_{1},\ldots,\mathfrak{K}_{\Theta}\right)/\mathfrak{K}{}_{\theta}$,
$\mathfrak{K}=\mathrm{lcm}\left(\mathfrak{K}_{1},\ldots,\mathfrak{K}_{\Theta}\right)$
and $\Psi_{\theta}^{C_{\theta}}\left(\pi_{\theta,1},\ldots,\pi_{\theta,n-r}\right)\in\left[1_{Q}\right]$.

Note that we can derive (\ref{eq:set2}) and (\ref{eq:set2b}) from
(\ref{eq:41}) by the method described here only if $\Phi$ is precomplete,
all $\Phi_{\theta}'$ are portable and all $\Phi_{\theta}^{*}$ have
scaling-covariant scalar representations, although we can obtain an
incomplete result if $\Phi$ is precomplete and there is at least
one $\theta$ such that $\Phi_{\theta}'$ is portable and $\Phi_{\theta}^{*}$
has a scaling-covariant scalar representation.

\subsection{Starting from a dimensional matrix}

The minimal exponents $K_{\theta,i},K_{\theta,ij}$ required in (\ref{eq:421})
depend on the structure of the subgroup $\left(Q/{\sim}\right)'$
of $Q/{\sim}$ generated by $\mathsf{C}_{0},\ldots,\mathsf{C}_{n}$.
As described below, it is usually most convenient to derive the structure
of $\left(Q/{\sim}\right)'$, and indirectly the minimal exponents,
from the structure of $Q/{\sim}$ by choosing a basis for $Q/{\sim}$.

Let $\Phi:\mathsf{C}_{1}\times\cdots\times\mathsf{C}_{n}\rightarrow\mathsf{C_{0}^{\mathfrak{K}}}$
be a precomplete quantity function on $Q$, and fix a basis $\mathsf{E}=\left\{ \mathsf{E}_{1},\ldots,\mathsf{E}_{m}\right\} $
for $Q/{\sim}$. Each $\mathsf{C}_{i}$ is associated with a unique
column vector $\varepsilon_{i}=\left(\varepsilon{}_{i1},\ldots,\varepsilon_{im}\right)^{\!\mathrm{T}}$
of integer exponents $\varepsilon_{i\ell}$ such that $\mathsf{C}_{i}=\prod_{\ell=1}^{m}\mathsf{E}_{\ell}^{\varepsilon_{i\ell}}$.
The \emph{dimensional matrix} for $Q$ and $\Phi$ relative to $\mathsf{E}$
is 
\begin{equation}
\begin{array}{ccccccc}
 & \mathsf{C}_{0} & \mathsf{C}_{1} & \cdots & \mathsf{C}_{i} & \cdots & \mathsf{C}_{n}\\
\mathsf{E}_{1} & \varepsilon_{01} & \varepsilon_{11} & \cdots & \varepsilon_{i1} & \cdots & \varepsilon_{n1}\\
\vdots & \vdots & \vdots &  & \vdots &  & \vdots\\
\mathsf{E}_{m} & \varepsilon_{0m} & \varepsilon_{1m} & \cdots & \varepsilon_{im} & \cdots & \varepsilon_{nm}
\end{array}.\label{eq:dm1}
\end{equation}

For $j=1,\ldots,s$, let $\overline{\varepsilon}_{j}=\left(\overline{\varepsilon}{}_{j1},\ldots,\overline{\varepsilon}{}_{jm}\right)^{\!\mathrm{T}}$
be the column vector associated with $\mathsf{\overline{C}}_{j}\in\left\{ \mathsf{C}_{0},\ldots,\mathsf{C}_{n}\right\} $.
It is clear from the construction of the dimensional matrix that each
$\overline{\mathsf{C}}_{j}$ is represented by $\overline{\varepsilon}_{j}$
relative to $\mathsf{E}$ in such a way that $\prod_{j=1}^{s}\mathsf{\overline{C}}_{j}^{k_{j}}=\left[1_{Q}\right]$
if and only if $\sum_{j=1}^{s}k_{j}\overline{\varepsilon}_{j}=\boldsymbol{0}^{m}$,
where $\boldsymbol{0}^{m}$ is a column vector with $m$ zeros. In
other words, $k_{1},\ldots,k_{s}$ satisfy the system of equations
\begin{equation}
\begin{cases}
\sum_{j=1}^{s}\nolimits\overline{\varepsilon}{}_{j\ell}k_{j}=0 & \qquad\left(\ell=1,\ldots,m\right)\end{cases}\label{eq:eq0}
\end{equation}

Hence, independence of dimensions is equivalent to linear independence
of columns in an associated dimensional matrix, so the rank of the
matrix is equal to the rank of the corresponding $\left(Q/{\sim}\right)'$
and to the number of elements in any prebasis of dimensions for $\Phi$
by Lemma \ref{lem:L21}.

Considering the functions $\Phi_{\theta}':\left(x_{\theta,1},\ldots,x_{\theta,r}\right)\left(y_{\theta,1},\ldots,y_{\theta,n-r}\right)\mapsto y_{0}^{\mathfrak{K_{\theta}}}$
in (\ref{eq:42}) it also follows from the equivalence of $\prod_{j=1}^{s}\mathsf{\overline{C}}_{j}^{k_{j}}=\left[1_{Q}\right]$
and equations (\ref{eq:eq0}) that, for $\theta=1,\ldots,\Theta$,
the sequences of minimal exponents $K_{\theta,i},K_{\theta,i1},\ldots,K_{\theta,ir}$
in (\ref{eq:421}) are obtained from the sets of sequences of integers
$k_{\theta,i},k_{\theta,i1},\ldots,k_{\theta,ir}$ that satisfy corresponding
systems of equations,
\begin{equation}
\begin{cases}
\eta_{0\ell}k_{\theta,0}=\sum_{j=1}^{r}\nolimits\xi{}_{\theta,j\ell}k_{\theta,0j} & \left(\ell=1,\ldots,m\right),\end{cases}\label{eq:eq1}
\end{equation}
\begin{equation}
\begin{cases}
\eta{}_{\theta,i\ell}k_{\theta,i}=\sum_{j=1}^{r}\nolimits\xi{}_{\theta,j\ell}k_{\theta,ij} & \left(\ell=1,\ldots,m\right)\qquad\left(i=1,\ldots,n-r\right),\end{cases}\label{eq:eq2}
\end{equation}
where $\left(\eta_{01},\ldots,\eta_{0m}\right)$, $\left(\eta{}_{\theta,i1},\ldots,\eta_{\theta,im}\right)$
and $\left(\xi_{\theta,j1},\ldots,\xi_{\theta,jm}\right)$ are the
$m$-tuples of exponents associated with $\left[y_{0}\right]$, $\left[y_{\theta,i}\right]$
and $\left[x_{\theta,j}\right]$, respectively, relative to $\mathsf{E}$.
\begin{example}
\label{ex2}Consider the equation $q_{0}^{\mathfrak{K}}=\Phi\left(q_{1},q_{2},q_{3}\right)$
and the corresponding dimensional matrices$\quad$
\[
\begin{array}{ccccc}
\left(A\right) & \left[q_{0}\right] & \left[q_{1}\right] & \left[q_{2}\right] & \left[q_{3}\right]\\
\mathsf{E}_{1} & 4 & 1 & 2 & 1\\
\mathsf{E}_{2} & 2 & 0 & 0 & 1
\end{array},
\]
\[
\begin{array}{ccccc}
\left(B\right) & \left[y_{0}\right] & \left[x_{1,1}\right] & \left[x_{1,2}\right] & \left[y_{1,1}\right]\\
\mathsf{E}_{1} & 4 & 1 & 1 & 2\\
\mathsf{E}_{2} & 2 & 0 & 1 & 0
\end{array},\qquad\begin{array}{ccccc}
\left(C\right) & \left[y_{0}\right] & \left[x_{2,1}\right] & \left[x_{2,2}\right] & \left[y_{2,1}\right]\\
\mathsf{E}_{1} & 4 & 2 & 1 & 1\\
\mathsf{E}_{2} & 2 & 0 & 1 & 0
\end{array},
\]
where $\left(B\right)$ and $\left(C\right)$ are obtained from $\left(A\right)$
by the prebasis reorderings $\left(q_{1},q_{2},q_{3}\right)\mapsto\left(q_{1},q_{3},q_{2}\right)=\left(x_{1,1},x_{1,2}\right)\left(y_{1,1}\right)$
and $\left(q_{1},q_{2},q_{3}\right)\mapsto\left(q_{2},q_{3},q_{1}\right)=\left(x_{2,1},x_{2,2}\right)\left(y_{2,1}\right)$,
respectively. From each one of $\left(B\right)$ and $\left(C\right)$,
we obtain two systems of two equations each,
\[
\begin{array}{cc}
\begin{cases}
4k_{1,0}=1k_{1,01}+1k_{1,02}\\
2k_{1,0}=0k_{1,01}+1k_{1,02}
\end{cases}\left(B_{0}\right),\qquad & \begin{cases}
2k_{1,1}=1k_{1,11}+1k_{1,12}\\
0k_{1,1}=0k_{1,11}+1k_{1,12}
\end{cases}\left(B_{1}\right),\end{array}
\]
\[
\begin{array}{cc}
\begin{cases}
4k_{2,0}=2k_{2,01}+1k_{2,02}\\
2k_{2,0}=0k_{2,01}+1k_{2,02}
\end{cases}\left(C_{0}\right),\qquad & \begin{cases}
1k_{2,1}=2k_{2,11}+1k_{2,12}\\
0k_{2,1}=0k_{2,11}+1k_{2,12}
\end{cases}\left(C_{1}\right).\end{array}
\]
The minimal exponents in (\ref{eq:421}), obtained by solving these
systems of equations, are
\begin{gather*}
\left(K_{1,0},K_{1,01},K_{1,02}\right)=\left(1,2,2\right)\quad\left(B_{0}\right),\qquad\left(K_{1,1},K_{1,11},K_{1,12}\right)=\left(1,2,0\right)\quad\left(B_{1}\right),\\
\left(K_{2,0},K_{2,01},K_{2,02}\right)=\left(1,1,2\right)\quad\left(C_{0}\right),\qquad\left(K_{2,1},K_{2,11},K_{2,12}\right)=\left(2,1,0\right)\quad\left(C_{1}\right).
\end{gather*}
Setting $\mathfrak{K}=K_{1,0}=K_{2,0}=1$, $q_{0}^{\mathfrak{K}}=\Phi\left(q_{1},q_{2},q_{3}\right)$
thus has the representations $y_{0}=x_{1,1}^{2}x_{1,2}^{2}\,\Psi_{1}\left(y_{1,1}/x_{1,1}^{2}\right)$
and $y_{0}=x_{2,1}x_{2,2}^{2}\,\Psi_{2}\left(y_{2,1}^{2}/x_{2,1}\right)$,
or alternatively $q_{0}=q_{1}^{2}q_{3}^{2}\,\Psi_{1}\left(q_{2}/q_{1}^{2}\right)$
and $q_{0}=q_{2}q_{3}^{2}\,\Psi_{2}\left(q_{1}^{2}/q_{2}\right)$,
provided that the required scaling-covariant scalar representations
exist and $\Phi_{2}'$ is portable although $K_{2,1}=2$.
\end{example}

\begin{rem}
It is known from the practice of dimensional analysis that a change
of the basis $\mathsf{E}$ for a dimensional matrix may lead to a
change of the minimal exponents associated with its variables. This
can only happen, however, if the change of basis is tied to a change
of quantity space. The conceptual reason for this is that a change
of basis for $Q/{\sim}$, where $Q$ is a fixed quantity space, does
not affect the dependencies in $Q/{\sim}$ or its subgroup $\left(Q/{\sim}\right)'$.
\end{rem}

\begin{example}
\label{ex43}Consider a (right, circular) cone. The slant height $H$
of the cone (that is, the length of the line segment from the periphery
of its base to its apex) is determined by the area of its base $a$
and its height $h$,
\[
H^{\mathfrak{K}}=\Phi\left(a,h\right).
\]
 The dimensional matrix for $\Phi$ is simply
\[
\begin{array}{cccc}
 & \left[H\right] & \left[a\right] & \left[h\right]\\
\mathsf{L} & 1 & 2 & 1
\end{array}.
\]
$\Phi$ is precomplete and has prebases of dimensions $\left\{ \left[a\right]\right\} $
and $\left\{ \left[h\right]\right\} $, so the system of equations
of the form (\ref{eq:42}) is
\[
\begin{cases}
H^{\mathfrak{K}_{1}}=\Phi_{1}'\left(a\right)\left(h\right) & \left(\mathrm{for}\;\left\{ \left[a\right]\right\} \right),\\
H^{\mathfrak{K}_{2}}=\Phi_{2}'\left(h\right)\left(a\right) & \left(\mathrm{for}\;\left\{ \left[h\right]\right\} \right),
\end{cases}
\]
and the minimal exponents in (\ref{eq:421}) are $\left(K_{1,0},K_{1,01}\right)=\left(2,1\right)$,
$\left(K_{1,1},K_{1,11}\right)=\left(2,1\right)$, $\left(K_{2,0},K_{2,01}\right)=\left(1,1\right)$
and $\left(K_{2,1},K_{2,11}\right)=\left(1,2\right)$. Thus, setting
$\mathfrak{K}_{1}=2$ and $\mathfrak{K}_{2}=1$ both equations become
consistent, and $\Phi_{2}$ is portable because $K_{2,1}=1$, so if
$\text{\ensuremath{h>0_{\left[h\right]}} so that }\Phi_{1}'$ is portable
as well then
\[
\begin{cases}
H^{2}=\Phi_{1}^{*}\left(a\right)\left(h^{2}\right),\\
H=\Phi_{2}^{*}\left(h\right)\left(a\right).
\end{cases}
\]
Hence, if $\Phi_{1}^{*}$ and $\Phi_{2}^{*}$ have scaling-covariant
scalar representations then we obtain
\[
\begin{cases}
H^{2}=a\Psi_{1}\left(h^{2}/a\right),\\
H=h\Psi_{2}\left(a/h^{2}\right),
\end{cases}
\]
and thus
\[
\begin{cases}
H^{2}=a\Psi_{1}\left(h^{2}/a\right), & \left(1\right)\\
H^{2}=h^{2}\Psi_{2}^{2}\left(a/h^{2}\right), & \left(2\right)
\end{cases}
\]
where $\Psi_{2}^{2}\left(a/h^{2}\right)=\Psi_{2}\left(a/h^{2}\right)^{2}$.
Note that $a=\pi r^{2}$, where $r$ is the radius of the circular
periphery of the base, so $\Phi\left(a,h\right)=H^{2}=r^{2}+h^{2}=a/\pi+h^{2}$.
The same function is obtained by substituting $\Psi_{1}\left(x\right)=1_{Q}/\pi+x$
in (1) or $\Psi_{2}^{2}\left(x\right)=1_{Q}+x/\pi$ in (2).
\end{example}

\section{\label{sec:5}Applications of dimensional analysis\label{sec:6}}

We start with two theoretically interesting examples of dimensional
analysis without dimensional matrices, and then move on to a little
more familiar applications with dimensional matrices as inputs. Only
quantity spaces over $\mathbb{R}$ will be considered in the examples
below.
\begin{example}
\label{x2-1}

By convention, $\prod_{j=1}^{0}e_{j}^{k_{j}}=1_{Q}$ and $\prod_{j=1}^{0}\mathsf{C}_{i}^{k_{j}}=\left[1_{Q}\right]$.
In particular, $\left[1_{Q}\right]=\prod_{j=1}^{0}\left[1_{Q}\right]^{k_{j}}$,
so $\emptyset$ is (vacuously) the unique prebasis of dimensions for
any quantity function of the form 
\[
\Phi:\left[1_{Q}\right]\times\cdots\times\left[1_{Q}\right]\rightarrow\left[1_{Q}\right],\qquad\left(q_{1},\ldots,q_{n}\right)\mapsto q_{0}.
\]
Thus, $\Theta=1$ and the sequences of minimal exponents $\left(K_{1,0},K_{1,0j}\right),\ldots,\left(K_{1,n},K_{1,nj}\right)$
associated with $\emptyset$ are all just the sequence $\left(1\right)$,
so $\Phi$ is complete with basis of dimensions $\emptyset$, and
we can apply Theorem \ref{pimain} directly. The unique expansion
of $\Phi\left(q_{1},\ldots,q_{n}\right)$ relative to $\emptyset$
is $\Phi\left(q_{1},\ldots,q_{n}\right)=\mu_{\emptyset}\left(\Phi\left(q_{1},\ldots,q_{n}\right)\right)\cdot1_{Q},$
so $\phi$ defined by 
\[
\phi\left(\mu_{\emptyset}\left(q_{1}\right),\ldots,\mu_{\emptyset}\left(q_{n}\right)\right)\cdot1_{Q}=\Phi\left(q_{1},\ldots,q_{n}\right)
\]
is a covariant scalar representation of $\Phi$ since $\phi\left(\mu_{\emptyset}\left(q_{1}\right),\ldots,\mu_{\emptyset}\left(q_{n}\right)\right)\cdot1_{Q}=\mu_{\emptyset}\left(\Phi\left(q_{1},\ldots,q_{n}\right)\right)\cdot1_{Q}$
and $1_{Q}$ is a unit quantity for $\left[1_{Q}\right]$. Hence,
\[
q_{0}=1_{Q}\Psi\left(q_{1}/1_{Q},\ldots,q_{n}/1_{Q}\right)=\Psi\left(q_{1},\ldots,q_{n}\right),
\]
so $\Phi=1_{Q}\Psi=\Psi$. We conclude that it is possible but pointless
to use dimensional analysis if all quantity variables range over $\left[1_{Q}\right]$.
\end{example}

\begin{example}
\label{x3-1}Consider a quantity function $\Phi$ of the form
\begin{equation}
\Phi:\mathsf{C}_{1}\rightarrow\left[1_{Q}\right],\qquad q_{1}\mapsto q_{0},\label{eq:const}
\end{equation}
where $\mathsf{C}_{1}\neq\left[1_{Q}\right]$, As $\left[1_{Q}\right]^{1}=\mathsf{C}_{1}^{0}$,
$\Phi$ is complete with basis of dimensions $\left\{ \mathsf{C}_{1}\right\} $,
and we can apply Theorem \ref{pimain} directly. 

Hence, if $\Phi$ has a scaling-covariant scalar representation then
\[
q_{0}=q_{1}^{0}\Psi\left(\right)=1_{Q}\Psi\left(\right)=k\in\left[1_{Q}\right],
\]
so $\Phi$ is a constant function. Had $\Phi$ not admitted a scaling-covariant
scalar representation then nothing could have been said about how
$\Phi\left(q_{1}\right)$ depends on $q_{1}$, so we obtain specific
information about $\Phi$ from this assumption. A non-constant function
of the form (\ref{eq:const}), complete but without a scaling-covariant
scalar representation, was defined in Example \ref{ex1}.
\end{example}

In the examples below, we assume that every $\Phi_{i}^{*}$ has a
scaling-covariant scalar representation, meaning that every $\Phi_{i}^{*}$
is ``physically meaningful''. 
\begin{example}
\label{x2}Assume that the period of oscillation $t$ of a pendulum
is determined by its length $\ell$, the mass of the bob $m$, the
amplitude of the oscillation $\theta$ (an angle) and the constant
of gravity $g$, that is,
\[
t^{\mathfrak{K}}=\Phi\left(\ell,m,\theta,g\right).
\]
The dependencies among the corresponding dimensions are given by the
dimensional matrix
\[
\begin{array}{cccccc}
 & \left[t\right] & \left[\ell\right] & \left[m\right] & \left[\theta\right] & \left[g\right]\\
\mathsf{L} & 0 & 1 & 0 & 0 & 1\\
\mathsf{T} & 1 & 0 & 0 & 0 & -2\\
\mathsf{M} & 0 & 0 & 1 & 0 & 0
\end{array}.
\]
The unique prebasis of dimensions for $\Phi$ is $\left\{ \left[\ell\right],\left[m\right],\left[g\right]\right\} $,
and from the minimal exponents calculated as in Example \ref{ex2}
we obtain
\[
\left[t\right]^{2}=\left[\ell\right]\left[m\right]^{0}\left[g\right]^{-1},\quad\left[\theta\right]=\left[\ell\right]^{0}\left[m\right]^{0}\left[g\right]^{0}.
\]
 Thus, 
\[
t^{2}=\Phi'\left(\ell,m,g\right)\left(\theta\right)
\]
is a consistent equation with a portable function, so we also have
\[
t^{2}=\Phi^{*}\left(\ell,m,g\right)\left(\theta\right),
\]
 and hence 
\begin{equation}
t^{2}=\ell m^{0}g^{-1}\Psi\left(\theta/\ell^{0}m^{0}g^{0}\right)=\ell g^{-1}\Psi\left(\theta\right).\label{eq:pendel}
\end{equation}

Consider a covariant scalar representation $\phi$ of $\Phi$ and
let $\boldsymbol{t}$, $\boldsymbol{\ell}$, $\boldsymbol{m}$, $\boldsymbol{g}$,
$\boldsymbol{g}^{-1}$ and $\boldsymbol{\theta}$ denote the scalars
$\mu_{E}\left(t\right)$, $\mu_{E}\left(\ell\right)$, $\mu_{E}\left(m\right)$,
$\mu_{E}\left(g\right)$, $\mu_{E}\left(g^{-1}\right)$ and $\mu_{E}\left(\theta\right)$,
respectively. Then $\boldsymbol{t}^{2}=\boldsymbol{\ell}\boldsymbol{g}^{-1}\psi\left(\boldsymbol{\theta}\right)$,
where $\psi$ is the covariant scalar representation of $\Psi:\left[1_{Q}\right]\rightarrow\left[1_{Q}\right]$
(recall Example \ref{x2-1}). If $\boldsymbol{\ell},\boldsymbol{g}>0$
so that $\boldsymbol{\ell g}^{-1}>0$ then $\psi\left(\boldsymbol{\theta}\right)\geq0$
since $\boldsymbol{t}^{2}\geq0$, so $\boldsymbol{t}=\sqrt{\boldsymbol{\ell}\boldsymbol{g}^{-1}}\sqrt{\psi\left(\boldsymbol{\theta}\right)}\geq0$.
If we require $\phi$ to be defined for $\theta=0$ and set $\boldsymbol{t}_{0}=\sqrt{\phi\left(\boldsymbol{\ell},\boldsymbol{m},0,\boldsymbol{g}\right)}=0$
then $\sqrt{\psi\left(0\right)}=0$ since $\sqrt{\phi\left(\boldsymbol{\ell},\boldsymbol{m},0,\boldsymbol{g}\right)}=\sqrt{\boldsymbol{\ell}\boldsymbol{g}^{-1}}\sqrt{\psi\left(0\right)}$,
but it is known that $\sqrt{\psi\left(\boldsymbol{\theta}\right)}\rightarrow2\pi$
as $\boldsymbol{\theta}\rightarrow0$ \cite{se}, so $\boldsymbol{t}\approx2\pi\sqrt{\boldsymbol{\ell}\boldsymbol{g}^{-1}}$
for small $\theta\neq0$. We note that the functions appearing in
dimensional analysis need not be continuous.
\end{example}

\begin{rem}
If $q\in\left[1_{Q}\right]$ then we may define $\sqrt[n]{q}$ as
$\sqrt[n]{\mu_{E}\left(q\right)}\cdot1_{Q}$ (if $\mu_{E}\left(q\right)\geq0)$,
$e^{q}$ as $e^{\mu_{E}\left(q\right)}\cdot1_{Q}$ and so on, since
$\mu_{E}\left(q\right)$ does not depend on a basis $E$. Conversely,
if $q\notin\left[1_{Q}\right]$ then $\sqrt[n]{q}\in Q$ only if $q=x^{n}$
for some $x\in Q$, where $n$ is a positive integer, since in a monoid
$x^{n}$ is defined in terms of repeated multiplication. In Example
\ref{x2}, $\Psi\left(\theta\right)\in\left[1_{Q}\right]$, $\mu_{E}\left(\Psi\left(\theta\right)\right)=\psi\left(\mu_{E}\left(\theta\right)\right)\geq0$
and $\ell g^{-1}=\mu_{E}\left(\ell g^{-1}\right)\cdot t^{2}=\left(\sqrt{\mu_{E}\left(\ell g^{-1}\right)}\cdot t\right)^{2}$
by Lemma \ref{lem:22} and since $\mu_{E}\left(\ell\right)\mu_{E}\left(g^{-1}\right)>0$,
so there are indeed quantities $\sqrt{\Psi\left(\theta\right)},\sqrt{\ell g^{-1}}\in Q$.
\end{rem}

\begin{example}
\label{x4}Let $\mathfrak{A}$ and $\mathfrak{B}$ be two bodies of
mass $a$ and $b$, respectively, let $c$ be the combined mass of
$\mathfrak{A}$ and $\mathfrak{B}$, and assume that we have 
\[
c^{\mathfrak{K}}=\Phi(a,b).
\]
The simple dimensional matrix is
\[
\begin{array}{cccc}
 & \left[c\right] & \left[a\right] & \left[b\right]\\
\mathsf{M} & 1 & 1 & 1
\end{array}.
\]
Thus, $\left\{ \left[a\right]\right\} $ and $\left\{ \left[b\right]\right\} $
are the two prebases of dimensions for $\Phi$, and from the minimal
exponents calculated from the two corresponding dimensional matrices
we obtain
\[
\begin{cases}
\left[c\right]=\left[a\right],\quad\left[b\right]=\left[a\right] & \left(\mathrm{for}\;\left\{ \left[a\right]\right\} \right),\\
\left[c\right]=\left[b\right],\quad\left[a\right]=\left[b\right] & \left(\mathrm{for}\;\left\{ \left[b\right]\right\} \right).
\end{cases}
\]
Hence, 
\[
\begin{cases}
c=\Phi_{1}'\left(a\right)\left(b\right)\\
c=\Phi_{2}'\left(b\right)\left(a\right)
\end{cases}
\]
are consistent equations with portable functions, so we have
\[
\begin{cases}
c=\Phi_{1}^{*}\left(a\right)\left(b\right)=a\,\Psi_{1}\left(b/a\right), & (M1)\\
c=\Phi_{2}^{*}\left(b\right)\left(a\right)=b\,\Psi_{2}\left(a/b\right). & (M2)
\end{cases}
\]

Thus, for $a,b\neq0_{\mathsf{M}}$ we have $\Phi\left(a,b\right)=a\,\Psi_{1}\left(b/a\right)$
by ($M1$) and $\Phi\left(b,a\right)=a\,\Psi_{2}\left(b/a\right)$
by ($M2$), so if we assume by symmetry that $\Phi(a,b)=\Phi(b,a)$
then $\Psi_{1}=\Psi_{2}=\Psi$. Hence, $a\,\Psi\left(b/a\right)=b\,\Psi\left(a/b\right)$,
so setting $x=b/a$ we obtain a functional equation of the form $\Psi\left(x\right)=F\left(x,\Psi\right)$,
namely
\[
\Psi\left(x\right)=x\,\Psi\left(x^{-1}\right).
\]
This equation has solutions of the form 
\[
\Psi(x)=k\left(1_{Q}+x\right)\qquad\left(k,x\in\left[1_{Q}\right]\right),
\]
unique under natural analyticity conditions (see Appendix A), so 
\[
c=a\,\Psi\left(b/a\right)=ak\left(1_{Q}+b/a\right)=k\left(a+b\right)\qquad\left(a\neq0_{\mathsf{M}}\right).
\]
If we assume that $\Phi(a,0_{\mathsf{M}})=a$ for all non-zero $a\in\mathsf{M}$
then $a=a\Psi(0_{\mathsf{\left[1_{Q}\right]}})=ak\left(1_{Q}+0_{\mathsf{\left[1_{Q}\right]}}\right)=ak$,
so $k=1_{Q}$, so we obtain $c=a+b$ as one might expect. If also
$\Phi(0_{\mathsf{M}},0_{\mathsf{M}})=0_{\mathsf{M}}$ then 
\[
c=a+b
\]
for all $a,b\in\mathsf{M}$ (taking care also of hypothetical negative
masses).
\end{example}

\begin{example}
\label{ex56}Consider a right triangle with sides $\mathfrak{A}$,
$\mathfrak{B}$ and $\mathfrak{C}$, where the right angle is the
one between $\mathfrak{A}$ and $\mathfrak{B}$. Let $a$, $b$ and
$c$ be the lengths of $\mathfrak{A}$, $\mathfrak{B}$ and $\mathfrak{C}$,
respectively, and $A$, $B$ and $C$ the areas of the squares whose
sides are $\mathfrak{A}$, $\mathfrak{B}$ and $\mathfrak{C}$, respectively.

We first want to find a function $\Phi$ such that $C=\Phi\left(A,B\right)$.
Assuming that $\Phi\left(A,B\right)=\Phi\left(B,A\right)$ and reasoning
as in Example \ref{x4}, we again derive the functional equation $\Psi\left(x\right)=x\,\Psi\left(x^{-1}\right)$.
Substituting the analytical solution $\Psi\left(x\right)=k\left(1_{Q}+x\right)$,
where $x=B/A$, in $\Phi\left(A,B\right)=A\Psi\left(B/A\right)$ we
obtain $\Phi\left(A,B\right)=k\left(A+B\right)$, and it is clear
that $\left(A+B\right)/C\rightarrow1_{Q}$ as $B\rightarrow0_{\left[B\right]}$,
so $k=1_{Q}$. Thus, we have $C=A+B$, Pythagoras' theorem.

In the same way, we could ``prove'' that $c=a+b$, but this is obviously
not true. Note, however, that $\Psi\left(x\right)=k\sqrt{1_{Q}+x^{2}}$,
where $x=b/a$, is also a solution of $\Psi\left(x\right)=x\,\Psi\left(x^{-1}\right)$,
albeit not an analytical one. Substituting this solution in $\Phi\left(a,b\right)=a\Psi\left(b/a\right)$
we obtain $\Phi\left(a,b\right)=k\left(\sqrt{a^{2}+b^{2}}\right)$,
and finally $c=\sqrt{a^{2}+b^{2}}$, Pythagoras' theorem in another
form.
\end{example}

This example shows that a solution $\Psi$ of the functional equation
cannot be discarded just because it is non-analytical. Thus, the non-uniqueness
of $\Psi$ may be a concern, but it is often obvious which $\Psi$
to choose. For example, in the right triangle case we obviously have
$c<a+b$. Presumably, in most cases of interest $\Psi$ is analytical,
and thus unique, and this will be assumed in the next two examples.

\begin{example}[based on a problem in Buckingham \cite{key-2}, pp. 358\textendash 359]
\label{x5}

It is assumed that the energy density $u$ at a certain point is determined
by the strengths $E$ and $H$ there of an electric field $\boldsymbol{\mathbf{E}}$
and a magnetic $\mathbf{H}$-field, respectively, as well as the permittivity
$\epsilon$ and permeability $\mu$ of the medium, that is, 
\begin{equation}
u^{\mathfrak{K}}=\Phi\left(E,H,\epsilon,\mu\right).\label{eq:61}
\end{equation}
The corresponding dimensional matrix is
\[
\begin{array}{cccccc}
 & \left[u\right] & \left[E\right] & \left[H\right] & \left[\epsilon\right] & \left[\mu\right]\\
\mathsf{L} & -1 & 1 & -1 & -3 & 1\\
\mathsf{T} & -2 & -3 & 0 & 4 & -2\\
\mathsf{M} & 1 & 1 & 0 & -1 & 1\\
\mathsf{I} & 0 & -1 & 1 & 2 & -2
\end{array}.
\]
It is easy to verify that the dimensional matrix has rank 3 and that
there are four prebases of dimensions for $\Phi$, namely $\left\{ \left[E\right],\left[\epsilon\right],\left[\mu\right]\right\} $,
$\left\{ \left[H\right],\left[\epsilon\right],\left[\mu\right]\right\} $,
$\left\{ \left[E\right],\left[H\right],\left[\epsilon\right]\right\} $
and $\left\{ \left[E\right],\left[H\right],\left[\mu\right]\right\} $.
From the minimal exponents we obtain
\[
\begin{cases}
\left[u\right]=\left[E\right]^{2}\left[\epsilon\right],\quad\left[H\right]^{2}=\left[E\right]^{2}\left[\epsilon\right]\left[\mu\right]^{-1} & \left(\mathrm{for}\;\left\{ \left[E\right],\left[\epsilon\right],\left[\mu\right]\right\} \right),\\
\left[u\right]=\left[H\right]^{2}\left[\mu\right],\quad\left[E\right]^{2}=\left[H\right]^{2}\left[\epsilon\right]^{-1}\left[\mu\right] & \left(\mathrm{for}\;\left\{ \left[H\right],\left[\epsilon\right],\left[\mu\right]\right\} \right),\\
\left[u\right]=\left[E\right]^{2}\left[\epsilon\right],\quad\left[\mu\right]=\left[E\right]^{2}\left[H\right]^{-2}\left[\epsilon\right], & \left(\mathrm{for}\;\left\{ \left[E\right],\left[H\right],\left[\epsilon\right]\right\} \right),\\
\left[u\right]=\left[H\right]^{2}\left[\mu\right],\quad\left[\epsilon\right]=\left[H\right]^{2}\left[\epsilon\right]^{-1}\left[\mu\right], & \left(\mathrm{for}\;\left\{ \left[E\right],\left[H\right],\left[\mu\right]\right\} \right),
\end{cases}
\]
giving the set of consistent equations, 
\[
\begin{cases}
u=\Phi_{1}'\left(E,\epsilon,\mu\right)\left(H\right),\\
u=\Phi_{2}'\left(H,\epsilon,\mu\right)\left(E\right),\\
u=\Phi_{3}'\left(E,H,\epsilon\right)\left(\mu\right),\\
u=\Phi_{4}'\left(E,H,\mu\right)\left(\epsilon\right),
\end{cases}
\]
where at least $\Phi_{3}'$ and $\Phi_{4}'$ are portable. If $\Phi_{1}'$
and $\Phi_{2}'$ are portable as well, we obtain
\begin{equation}
\begin{cases}
u=\Phi_{1}^{*}\left(E,\epsilon,\mu\right)\left(H^{2}\right)=E^{2}\epsilon\,\Psi_{1}\left(H^{2}/\left(E^{2}\epsilon\mu^{-1}\right)\right),\\
u=\Phi_{2}^{*}\left(H,\epsilon,\mu\right)\left(E^{2}\right)=H^{2}\mu\,\Psi_{2}\left(E^{2}/\left(H^{2}\epsilon^{-1}\mu\right)\right),\\
u=\Phi_{3}^{*}\left(E,H,\epsilon\right)\left(\mu\right)=E^{2}\epsilon\,\Psi_{3}\left(\mu/\left(E^{2}H^{-2}\epsilon\right)\right),\\
u=\Phi_{4}^{*}\left(E,H,\mu\right)\left(\epsilon\right)=H^{2}\mu\,\Psi_{4}\left(\epsilon/\left(E^{-2}H^{2}\mu\right)\right).
\end{cases}\label{eq:rep61}
\end{equation}
We note that $E,H,\epsilon,\mu$ occur only in the combinations $E'=\epsilon E^{2}$
and $H'=\mu H^{2}$ in these equations, so we can set $\varPhi\left(E',H'\right)=\Phi\left(E,H,\epsilon,\mu\right)$.
Also, $\Psi_{1}=\Psi_{3}$ and $\Psi_{2}=\Psi_{4}$, so it suffices
to consider $\Psi_{3}$ and $\Psi_{4}$, corresponding to the manifestly
portable functions $\Phi_{3}'$ and $\Phi_{4}'$. Thus, we can reduce
(\ref{eq:rep61}) to
\begin{equation}
\begin{cases}
u=E'\,\varPsi_{3}\left(H'/E'\right), & (EM1),\\
u=H'\,\varPsi_{4}\left(E'/H'\right). & (EM2).
\end{cases}\label{eq:buckrep}
\end{equation}

Although he considers only the sets $\left\{ E,\epsilon,\mu\right\} $
and $\left\{ H,\epsilon,\mu\right\} $ of ``fundamental'' quantities,
Buckingham, too, finds the representations in (\ref{eq:buckrep}),
writing them as $u=\epsilon E^{2}\,\varphi_{1}\left(\frac{\mu H^{2}}{\epsilon E^{2}}\right)$
and $u=\mu H^{2}\,\varphi_{2}\left(\frac{\epsilon E^{2}}{\mu H^{2}}\right)$,
respectively \cite[p. 359]{key-2}. He then remarks:
\begin{quotation}
\noindent {\small{}Assuming that the complete formula is $u=\frac{1}{8\pi}\left(\epsilon E^{2}+\mu H^{2}\right)$
we have $\varphi_{1}\left(x\right)=\varphi_{2}\left(x\right)=\frac{1+x}{8\pi}.$}{\small \par}
\end{quotation}
However, the purpose of dimensional analysis in this case is not to
derive $\varphi_{1}$ and $\varphi_{2}$ from the sought-after function
$\phi$ such that $u=\phi\left(E',H'\right)$ but to do the opposite,
so let us reverse Buckingham's inference, proceeding as in Example
\ref{x4}.

We have $\left[u\right]=\left[E'\right]=\left[H'\right]\neq\left[1_{Q}\right]$
and if we assume for symmetry reasons that $u=\varPhi\left(E',H'\right)=\varPhi\left(H',E'\right)$
and set $x=H'/E'$ then we obtain the functional equation $\varPsi\left(x\right)=x\,\varPsi\left(x^{-1}\right)$
as before. Thus, $\varPsi_{3}\left(x\right)=\varPsi_{4}\left(x\right)=\varPsi\left(x\right)=k\left(1_{Q}+x\right)$,
where $k,x\in\left[1_{Q}\right]$, and substitution in $\left(EM1\right)$
or $\left(EM2\right)$ gives
\[
u=k\left(E'+H'\right)=k\left(\epsilon E^{2}+\mu H^{2}\right),
\]
or $u=k\left(\epsilon E^{2}+\mu^{-1}B^{2}\right)$ if $B=\mu H$.

Working with examples, Buckingham thus recognised in \cite{key-2}
that there may be more than one way of representing $\Phi$. However,
he dismissed this observation by asserting that the representation
with $\varphi_{1}$ is ``equivalent'' to that with $\varphi_{2}$
\cite[p. 359, 362]{key-2}. While this equivalence is not obvious
from Buckingham's presentation, $\varPsi_{3}$ $\left(\phi_{1}\right)$
and $\varPsi_{4}$ $\left(\phi_{2}\right)$ are in fact equal, but
on the assumption that $\varPhi\left(E',H'\right)=\varPhi\left(H',E'\right)$.
\end{example}

\begin{example}[based on a problem in Bridgman \cite{bri}, pp.~5\textendash 8]
\label{x6} Let two bodies $\mathfrak{B}$ and $\mathfrak{b}$ with
masses $M$ and $m$ revolve around each other in empty space under
influence of their mutual gravitational attraction, as in the classical
two-body problem. Let $t$ denote the period of revolution and $d$
the mean distance between $\mathfrak{B}$ and $\mathfrak{b}$ (or
another characteristic distance). One might want to find out how $t$
depends on $M$, $m$ and $d$, \textcolor{black}{that is,
\[
t^{\mathfrak{K}}=\Phi_{0}\left(M,m,d\right),
\]
but the associated dimensional matrix
\[
\begin{array}{ccccc}
 & \left[t\right] & \left[M\right] & \left[m\right] & \left[d\right]\\
\mathsf{L} & 0 & 0 & 0 & 1\\
\mathsf{T} & 1 & 0 & 0 & 0\\
\mathsf{M} & 0 & 1 & 1 & 0
\end{array}
\]
shows that $\Phi_{0}$ is not precomplete; there is no prebasis of
dimensions for $\Phi_{0}$. }

\textcolor{black}{Bridgman suggests that $t$ does also depend on
the gravitational constant }\textcolor{black}{\emph{G}}\textcolor{black}{,
that is, 
\[
t^{\mathfrak{K}}=\Phi\left(M,m,d,G\right).
\]
giving the dimensional matrix}
\[
\begin{array}{cccccc}
 & \left[t\right] & \left[M\right] & \left[m\right] & \left[d\right] & \left[G\right]\\
\mathsf{L} & 0 & 0 & 0 & 1 & 3\\
\mathsf{T} & 1 & 0 & 0 & 0 & -2\\
\mathsf{M} & 0 & 1 & 1 & 0 & -1
\end{array}.
\]
Then $\left\{ \left[M\right],\left[d\right],\left[G\right]\right\} $
and $\left\{ \left[m\right],\left[d\right],\left[G\right]\right\} $
are the two prebases of dimensions for $\Phi$, and the minimal exponents
yield
\[
\begin{cases}
\left[t\right]^{2}=\left[M\right]^{-1}\left[d\right]^{3}\left[G\right]^{-1},\quad\left[m\right]=\left[M\right] & \left(\mathrm{for}\;\left\{ \left[M\right],\left[d\right],\left[G\right]\right\} \right),\\
\left[t\right]^{2}=\left[m\right]^{-1}\left[d\right]^{3}\left[G\right]^{-1},\quad\left[M\right]=\left[m\right] & \left(\mathrm{for}\;\left\{ \left[m\right],\left[d\right],\left[G\right]\right\} \right),
\end{cases}
\]
so
\[
\begin{cases}
t^{2}=\Phi_{1}'\left(M,d,g\right)\left(m\right)\\
t^{2}=\Phi_{2}'\left(m,d,g\right)\left(M\right)
\end{cases}
\]
are consistent equations with portable functions. Hence, we have
\[
\begin{cases}
t^{2}=\Phi_{1}^{*}\left(M,d,G\right)\left(m\right)=M^{-1}d^{3}G^{-1}\,\Psi_{1}\left(m/M\right), & (K1)\\
t^{2}=\Phi_{2}^{*}\left(m,d,G\right)\left(M\right)=m^{-1}d^{3}G^{-1}\,\Psi_{2}\left(M/m\right). & (K2)
\end{cases}
\]

Combining $\left(K1\right)$ and $\left(K2\right)$ we obtain $M^{-1}\Psi_{1}\left(m/M\right)=m^{-1}\Psi_{2}\left(M/m\right)$,
and assuming for symmetry reasons that $\Phi\left(M,m,d,G\right)=\Phi\left(m,M,d,G\right)$
this gives $M^{-1}\Psi_{1}\left(m/M\right)=M^{-1}\Psi_{2}\left(m/M\right)$.
This implies that $\Psi_{1}=\Psi_{2}=\Psi$, so setting $x=m/M$ we
obtain the functional equation
\[
\Psi\left(x\right)=x^{-1}\,\Psi\left(x^{-1}\right).
\]
This functional equation has solutions of the form 
\[
\Psi(x)=k\left(1_{Q}+x\right)^{-1}\qquad\left(k,x\in\left[1_{Q}\right]\right),
\]
unique under natural analyticity conditions (see Appendix A). Substituting
this in ($K1$) or ($K2$) gives
\begin{equation}
t^{2}=kd^{3}G^{-1}(M+m)^{-1}.\label{eq:kepler}
\end{equation}

Here, $k$ and $G$ are constants, so if several planets $\mathfrak{b}_{i}$
orbit the sun and $M+m_{i}\approx M$ for every planet mass $m_{i}$
then, approximately, $t^{2}\propto d^{3}$ regardless of relative
planetary masses. This is Kepler's third law of planetary motion.
(The mutual gravitational attraction of planets can be disregarded
since $m_{i}\ll M$.)

As before, (\ref{eq:kepler}) can be interpreted both as a quantity
equation and as a scalar equation, and assuming that $t,d,G,M,m>0$
the latter can also be written as 
\begin{equation}
t=k\sqrt{d^{3}G^{-1}(M+m)^{-1}}\qquad\left(k\in\mathbb{R}\right).\label{eq:scalOrb}
\end{equation}

It is worth pointing out that Bridgman \cite[p. 8]{bri} considered
only one equation, namely 
\[
t=\nicefrac{r^{\frac{3}{2}}}{\,G^{\frac{1}{2}}m_{2}^{\frac{1}{2}}}\,\phi\left(\frac{m_{2}}{m_{1}}\right).
\]
This corresponds to ($K1$) with $r=d$, $m_{1}=m$, $m_{2}=M$ and
$\phi\left(x\right)^{2}=\Psi_{1}\left(x^{-1}\right)$. Unlike Buckingham
before him, Bridgman did not reflect on the possibility that the original
function could have more than one representation, and apparently none
of them reflected on whether it was possible to obtain a stronger
result by combining dimensional analysis with a symmetry assumption.
\end{example}

\begin{rem}
The symmetry assumptions in Examples \ref{x4} \textendash{} \ref{x6}
can be applied to $\Psi_{1}$ and $\Psi_{2}$ separately. In Example
\ref{x4}, for example. if $\Phi\left(a,b\right)=a\,\Psi_{1}\left(b/a\right)$
and $\Phi\left(a,b\right)=\Phi\left(b,a\right)$ then $a\,\Psi_{1}\left(b/a\right)=b\,\Psi_{1}\left(a/b\right)$,
and we obtain the functional equation $\Psi_{1}\left(x\right)=x\,\Psi_{1}\left(x^{-1}\right)$
with the solutions $\Psi_{1}(x)=k_{1}\left(1_{Q}+x\right)$. Similarly,
combining $\Psi_{2}$ with $\Phi\left(a,b\right)=\Phi\left(b,a\right)$
we obtain $\Psi_{2}(x)=k_{2}\left(1_{Q}+x\right)$. However, recall
that the symmetry assumption also allows us to conclude that $\Psi_{1}=\Psi_{2}$,
implying that $k_{1}=k_{2}$.

Conversely, we may heuristically look for a symmetry of $\Phi$ that
makes it possible to conclude that $\Psi_{1}=\Psi_{2}$. For example,
in Example \ref{x6} it is not difficult to see that the symmetry
$\Phi\left(M,m,d,G\right)=\Phi\left(m,M,d,G\right)$ leads to the
conclusion that $\Psi_{1}=\Psi_{2}$. The symmetry of $\Phi$ can
in turn be derived from an assumption that ``all motion is relative''
so that a description of the two-body system in which $\mathfrak{B}$
is at rest and $\mathfrak{b}$ revolves around $\mathfrak{B}$ is
equivalent to a description in which $\mathfrak{b}$ is at rest and
$\mathfrak{B}$ revolves around $\mathfrak{b}$. Thus, we derive (\ref{eq:kepler}),
an empirically falsifiable generalization, from a relativity principle
together with the assumption that $M,m,d,G$ determine $t$ and assumptions
about the dimensions of these variables. By falsifying (\ref{eq:kepler})
we would falsify at least one postulate from which it was derived,
but by verifying (\ref{eq:kepler}) we provide support for the postulates,
including a relative motion assumption implying the symmetry of $\Phi$.
\end{rem}

\section{\label{sec:6-1}Dimensional analysis and matroid theory}

Recall that a (finite) matroid is a finite set $\mathcal{E}$ equipped
with a set $\mathcal{I}$ of subsets of $\mathcal{E}$, said to be
\emph{independent} sets. $\mathcal{I}$ is required to satisfy certain
conditions so as to generalize the notion of (linear) independence
of columns in a matrix, vectors in a vector space, elements of a free
abelian group etc. Those subsets of $\mathcal{E}$ which are not independent
are said to be \emph{dependent}. In matroid theory, a maximal independent
subset of $\mathcal{E}$ is called a \emph{basis}, and a minimal dependent
subset of $\mathcal{E}$ is called a \emph{circuit}. A subset $C$
of $\mathcal{E}$ such that $C\setminus\left\{ e\right\} $ is a basis
for some $e\in C$ is called a \emph{pseudocircuit} in this article.
Every matroid basis contains the same number $r$ of elements of $\mathcal{E}$,
so every pseudocircuit is a set of $r+1$ elements with a subset that
is a basis.

An integer matrix $D$ with labelled columns
\[
\begin{array}{ccc}
t_{1} & \cdots & t_{p}\\
\varepsilon_{11} & \cdots & \varepsilon_{1p}\\
\vdots & \ddots & \vdots\\
\varepsilon_{m1} & \cdots & \varepsilon_{mp}
\end{array}
\]
can be interpreted as a matroid with $\mathcal{E}=\left\{ t_{1},\ldots,t_{p}\right\} $
and the\emph{ }independent subsets of $\mathcal{E}$ corresponding
to the sets of linearly independent associated columns in $D$. A
maximal independent subset of $\mathcal{E}$ corresponds to a maximal
independent set of columns in $D$, so $r$ is equal to the rank of
$D$.

$D$ can also be interpreted in several ways as a dimensional matrix
associated with a quantity space $Q$, a quantity function on $Q$
and a basis for $Q/$$\!\sim$,
\[
\begin{array}{cccc}
 & \left[t_{1}\right] & \cdots & \left[t_{p}\right]\\
\mathsf{E}_{1} & \varepsilon_{11} & \cdots & \varepsilon_{1p}\\
 & \vdots & \ddots & \vdots\\
\mathsf{E}_{m} & \varepsilon_{m1} & \cdots & \varepsilon_{mp}
\end{array}.
\]
As elaborated below, a matroid basis corresponds to a prebasis for
a quantity function in augmented dimensional analysis and to a group
of ``repeating variables\emph{''} in certain formulations of dimensional
analysis (see, e.g., \cite{whi,key-11}), while a pseudocircuit corresponds
to a $\pi$-monomial in augmented dimensional analysis and a certain
kind of ``dimensionless group'' or \emph{``}$\pi$-group\emph{''}
in classical dimensional analysis.

\begin{example}
Consider the set $\left\{ a,b,c\right\} $ with the matroid structure
given by the matrix
\[
\begin{array}{ccc}
a & b & c\\
1 & 1 & 1\\
0 & 1 & 1
\end{array}.
\]
Here, $\emptyset$, $\left\{ a\right\} $, $\left\{ b\right\} $,
$\left\{ c\right\} $, $\left\{ a,b\right\} $ and $\left\{ a,c\right\} $
are independent sets, $\left\{ a,b\right\} $ and $\left\{ a,c\right\} $
are bases, $\left\{ a,b,c\right\} $ is a pseudocircuit but not a
circuit, and $\left\{ b,c\right\} $ is a circuit but not a pseudocircuit.
\end{example}

For every pseudocircuit $\left\{ s_{1},\ldots,s_{r+1}\right\} \subseteq\left\{ t_{1},\ldots,t_{p}\right\} $,
there are integers $K_{j}$, unique up to sign, that define a pair
of $\pi$\emph{-monomials} of the form $s_{1}^{K_{1}}\cdots s_{r+1}^{K_{r+1}}$
such that 
\begin{gather}
\prod_{j=1}^{r+1}\nolimits\left[s_{j}\right]^{K_{j}}=\left[1_{Q}\right],\qquad\mathrm{or\;equivalently}\qquad\sum\nolimits _{j=1}^{r+1}K_{j}\varepsilon_{\ell j}=0\quad\left(\ell=1,\ldots,m\right),\label{eq:cpi}\\
\exists j:K_{j}\neq0,\qquad\mathrm{gcd}\left(K_{1},\ldots,K_{r+1}\right)=1.
\end{gather}
The uniqueness of the integers $K_{j}$ follows from the rank-nullity
theorem for free $\mathbb{Z}$-modules, according to which the kernel
of an $m\times\left(r+1\right)$ integer matrix of rank $r$ has rank
1; see also \cite{jo2}.

Note that a monomial $t_{1}^{K_{1}}\cdots t_{p}^{K_{p}}$ satisfying
$\prod_{j=1}^{p}\left[t_{j}\right]^{K_{j}}=\left[1_{Q}\right]$ instead
of (\ref{eq:cpi}) need not be a $\pi$-monomial, and that the proof
of Theorem \ref{pimain} requires that the variables $\pi_{0},\ldots,\pi_{n-r}$
are $\pi$-monomials, in that context written as $y_{i}^{K_{i}}\prod_{j=1}^{r}\nolimits t_{j}^{-K_{ij}}$
with $K_{i}>0$.
\begin{example}
\label{ex62}For the matrix 
\[
\begin{array}{ccc}
a & b & c\\
1 & 2 & 1
\end{array},
\]
there are three pseudocircuits $\left\{ a,b\right\} ,\left\{ a,c\right\} ,\left\{ b,c\right\} $
with associated $\pi$-monomial pairs $\left\{ a^{2}b^{-1},a^{-2}b\right\} $,
$\left\{ ac^{-1},a^{-1}c\right\} $ and $\left\{ bc^{-2},b^{-1}c^{2}\right\} $.
Although $ab^{-1}c,a^{-1}bc^{-1}\in\left[1_{Q}\right]$ as well, these
invariant monomials or ``dimensionless groups'' are not appropriate
for dimensional analysis. In this case, augmented dimensional analysis
yields, for example, the equation $a=c\Psi\left(b/c^{2}\right)$,
but not $a=kbc^{-1}$ as obtained by setting $\Psi\left(x\right)=kx$.
Only if $\left\{ a,b,c\right\} $ is a pseudocircuit does the fact
that $ab^{-1}c\in\left[1_{Q}\right]$ for quantity variables $a,b,c$
imply that there is a constant $k$ such that $ab^{-1}c=k\in\left[1_{Q}\right]$
(compare Example \ref{ex43}).

Furthermore, $t_{1}^{K_{1}}\cdots t_{p}^{K_{p}}$ can be identified
with an element $\left(K_{1},\ldots,K_{p}\right)$ of $\mathbb{Z}^{p}$
as a free $\mathbb{Z}$-module, making the set of monomials satisfying
$\prod_{j=1}^{p}\left[t_{j}\right]^{K_{j}}=\left[1_{Q}\right]$ a
free submodule $\mathcal{N}$ of $\mathbb{Z}^{p}$, that is, a free
abelian group. It can be shown that a prebasis (maximal independent
subset) of $\mathcal{N}$ can be formed by selecting $p-r$ $\pi$-monomials
from as many $\pi$-monomial pairs. For example, $ab^{-1}c=\left(a^{2}b^{-1}\right)\left(ac^{-1}\right)^{-1}=\left(ac^{-1}\right)\left(b^{-1}c^{2}\right)$
and $\left(ab^{-1}c\right)^{2}=\left(a^{2}b^{-1}\right)\left(b^{-1}c^{2}\right)$,
with unique minimal exponents, so $\left\{ a^{2}b^{-1},ac^{-1}\right\} $,
$\left\{ ac^{-1},b^{-1}c^{2}\right\} $ and $\text{\ensuremath{\left\{  a^{2}b^{-1},b^{-1}c^{2}\right\} } are prebases for \ensuremath{\mathcal{N}}}$.
Thus, $ab^{-1}c$ and $a^{-1}bc^{-1}$ are redundant as well as inappropriate
for dimensional analysis, as are all invariant monomials except $\pi$-monomials.

\end{example}

\begin{example}
\label{ex9}Recall that the structure of a matroid is fully specified
by its set of bases. This set, and the set of pseudocircuits, can
be exhibited in a compact form as a matrix. Such a matrix is shown
below together with the dimensional matrix in Example \ref{x6} that
fully characterizes the matroid.
\begin{equation}
\begin{array}{cccccc}
 & \left[t\right] & \left[M\right] & \left[m\right] & \left[d\right] & \left[G\right]\\
\mathsf{L} & 0 & 0 & 0 & 1 & 3\\
\mathsf{T} & 1 & 0 & 0 & 0 & -2\\
\mathsf{M} & 0 & 1 & 1 & 0 & -1
\end{array},\qquad\begin{array}{cccccc}
 & t & M & m & d & G\\
A & + & + & - & + & -\\
B & + & + & - & - & +\\
C & + & - & \text{+} & + & -\\
D & + & - & + & - & +\\
E & + & - & - & + & +\\
F & - & + & - & + & +\\
G & - & - & + & + & +\\
\alpha & \ast & \ast & \ast & \ast & \circ\\
\beta & \ast & \ast & \ast & \circ & \ast\\
\gamma & \ast & \ast & \circ & \ast & \ast\\
\delta & \ast & \circ & \ast & \ast & \ast\\
\epsilon & \circ & \ast & \ast & \ast & \ast
\end{array}.\label{eq:matmat}
\end{equation}

Each row labelled by a capital letter specifies a matroid basis that
contains variables with a plus sign in this row, and each row labelled
by a Greek letter specifies a pseudocircuit that contains variables
with an asterisk in this row. For example, row $A$ in the table above
describes the basis $\left\{ t,M,d\right\} $, contained in the pseudocircuits
$\alpha=\left\{ t,M,m,d\right\} $ and $\gamma=\left\{ t,M,d,G\right\} $.

Corresponding to the pseudocircuits $\alpha$, $\beta$, $\mathfrak{\gamma}$,
$\mathfrak{\delta}$ and $\epsilon$, we have the $\pi$\emph{-}monomial
pairs 
\[
\pi_{\alpha}^{\pm1}=\pi_{\beta}^{\pm\text{1 }}=\pi_{\epsilon}^{\pm1}=\left\{ \pi_{a},\pi_{a}^{-1}\right\} ,\quad\pi_{\gamma}^{\pm1}=\left\{ \pi_{\gamma},\pi_{\gamma}^{-1}\right\} ,\quad\pi_{\delta}^{\pm1}=\left\{ \pi_{\delta},\pi_{\delta}^{-1}\right\} ,
\]
where we may omit variables with zero exponents and choose signs of
exponents so that
\[
\pi_{\alpha}=\pi_{\beta}=\pi_{\epsilon}=Mm^{-1},\quad\pi_{\gamma}=t^{2}Md^{-3}G,\quad\pi_{\delta}=t^{2}md^{-3}G.
\]

The $\pi$-monomials are the building blocks of equations of the form
$\pi_{0}=\Psi\left(\pi_{1},\ldots,\pi_{n-r}\right)$, but there are
constraints on which $\pi$-monomials that may appear in the same
equation. Recalling the discussion in Section \ref{sec:3}, we realize
that if $\pi_{0}=\Psi\left(\pi_{1},\ldots,\pi_{n-r}\right)$ represents
$y_{0}^{\mathfrak{K}}=\Phi^{*}\left(x_{1},\ldots,x_{r}\right)\left(y_{1},\ldots,y_{n-r}\right)$
then every $\pi$-monomial $\pi_{0},\ldots,\pi_{n}$ may be written
as
\begin{equation}
\pi_{i}=y_{i}^{K_{i}}\prod_{j=1}^{r}\nolimits x_{j}^{-K_{ij}},\label{eq:pimon}
\end{equation}
where $K_{i}\neq0$ and $\left\{ x_{1},\ldots,x_{r}\right\} $ is
a basis for $\Phi^{*}$, Thus, the unique basis for $\Phi^{*}$ is
the intersection of all pseudocircuits corresponding to $\pi$-monomials
in $\pi_{0}=\Psi\left(\pi_{1},\ldots,\pi_{n-r}\right)$. For example,
the combinations $\left\{ \pi_{\alpha}^{\pm1},\pi_{\gamma}^{\pm1}\right\} $,
$\left\{ \pi_{\alpha}^{\pm1},\pi_{\delta}^{\pm1}\right\} $ and $\left\{ \pi_{\gamma}^{\pm1},\pi_{\delta}^{\pm1}\right\} $
of $\pi_{\alpha}^{\pm1}$, $\pi_{\gamma}^{\pm1}$ and $\pi_{\delta}^{\pm1}$
are all legitimate since $\alpha\cap\gamma=A$, $\alpha\cap\delta=C$
and $\gamma\cap\delta=E$, respectively. 

It is also clear that any two equations $\pi_{0}=\Psi\left(\pi_{1},\ldots,\pi_{n-r}\right)$
and $\pi_{0}'=\Psi'\left(\pi_{1}',\ldots,\pi_{n-r}'\right)$ such
that $\pi_{i}$ and $\pi_{i}'$ belong to the same $\pi$-monomial
pair for each $i$ are equivalent, so it suffices to include one $\pi$-monomial
from each pair in the equations. 

Note that $\varPi=\left\{ \pi_{0},\pi_{1},\ldots,\pi_{n-r}\right\} $
is a set of $n+1-r$ independent monomials in $\mathcal{N}$, so $\varPi$
is a prebasis for $\mathcal{N}$ since $\mathcal{N}$ has rank $n+1-r$.
In the case considered here, $\mathcal{N}$ has rank $5-3=2$, and
$\left\{ Mm^{-1},t^{2}Md^{-3}G\right\} $, $\left\{ Mm^{-1},t^{2}md^{-3}G\right\} $
and $\left\{ t^{2}Md^{-3}G,t^{2}md^{-3}G\right\} $ are prebases for
$\mathcal{N}$ as well as sets of $\pi$-monomials associated with
the same matroid basis through pseudocircuits.

Rewriting $\pi_{0}=\Psi_{i}\left(\pi_{1},\ldots,\pi_{n-r}\right)$
as $y_{0}^{K_{0}}=\prod_{j=1}^{r}x_{j}^{K_{0j}}\Psi_{i}\left(\pi_{1},\ldots,\pi_{n-r}\right)$,
where $K_{0}>0$, we obtain five systems of quantity equations with
associated bases.

\[
\begin{cases}
t^{2}=M^{-1}d^{3}G^{-1}\,\Psi_{11}\left(m/M\right), & \left(F\right)\\
t^{2}=m^{-1}d^{3}G^{-1}\,\Psi_{12}\left(M/m\right), & \left(G\right)
\end{cases}
\]
\[
\begin{cases}
d^{3}=t^{2}MG\Psi_{21}\left(m/M\right), & \left(B\right)\\
d^{3}=t^{2}mG\Psi_{22}\left(M/m\right), & \left(D\right)
\end{cases}\qquad\begin{cases}
G=t^{-2}M^{-1}d^{3}\Psi_{31}\left(m/M\right), & \left(A\right)\\
G=t^{-2}m^{-1}d^{3}\Psi_{32}\left(M/m\right), & \left(C\right)
\end{cases}
\]
\[
\begin{cases}
M=m\Psi_{41}\left(G/t^{-2}m^{-1}d^{3}\right), & \left(C\right)\\
M=m\Psi_{42}\left(t^{2}/m^{-1}d^{3}G^{-1}\right), & \left(G\right)\\
M=m\Psi_{43}\left(d^{3}/t^{2}mG\right), & \left(D\right)\\
M=t^{-2}d^{3}G^{-1}\Psi_{44}\left(m/t^{-2}d^{3}G^{-1}\right), & \left(E\right)
\end{cases}\qquad\begin{cases}
m=M\Psi_{51}\left(G/t^{-2}M^{-1}d^{3}\right), & \left(A\right)\\
m=M\Psi_{52}\left(t^{2}/M^{-1}d^{3}G^{-1}\right), & \left(F\right)\\
m=M\Psi_{53}\left(d^{3}/t^{2}MG\right), & \left(B\right)\\
m=t^{-2}d^{3}G^{-1}\Psi_{54}\left(M/t^{-2}d^{3}G^{-1}\right). & \left(E\right)
\end{cases}
\]

We recognize the first system of equations as that already derived
in Example \ref{x6}. 

Note that $\Psi_{41}=\Psi_{42}$ since we have $G/t^{-2}m^{-1}d^{3}=t^{2}/m^{-1}d^{3}G^{-1}$,
and $\Phi_{42}\left(x\right)=\Phi_{43}\left(x^{-1}\right)$ since
$t^{2}/m^{-1}d^{3}G^{-1}=t^{2}mG/d^{3}$. This is because the corresponding
equations have the form 
\[
\pi_{\alpha}^{\pm1}=\Psi\left(\pi_{\delta}^{\pm1}\right)\qquad\left(\pi_{\alpha}^{\pm1}=\pi_{\beta}^{\pm\text{1 }}=\pi_{\epsilon}^{\pm1}\right)
\]
and $C\cup G\cup D=\delta$ whereas $\alpha\setminus\left\{ M\right\} =C$,
$\beta\setminus\left\{ M\right\} =D$ and $\epsilon\setminus\left\{ M\right\} =G$.
Hence, one can decide which one of the equivalent functions to keep
by choosing one of the bases contained in $\delta$. 

Similarly, the first three equations in the last system of equations
have the form 
\[
\pi_{\alpha}^{\pm1}=\Psi\left(\pi_{\gamma}^{\pm1}\right)\qquad\left(\pi_{\alpha}^{\pm1}=\pi_{\beta}^{\pm\text{1 }}=\pi_{\epsilon}^{\pm1}\right)
\]
and furthermore $A\cup F\cup B=\gamma$ whereas $\alpha\setminus\left\{ m\right\} =A$,
$\beta\setminus\left\{ m\right\} =B$ and $\epsilon\setminus\left\{ m\right\} =F$,
so one can decide which one of the equivalent functions $\Psi_{51}$,
$\Psi_{52}$ and $\Psi_{53}$ to keep by choosing one of $A$, $F$
and $B$.
\end{example}

It should be emphasized that the equations in Example \ref{ex9} were
obtained through formal manipulation of data in an integer matrix.
There is no guarantee that an equation constructed in this way is
mathematically well-behaved, makes sense conceptually and is of interest.
On the other hand, the ``balanced'' form of dimensional analysis
described in Example \ref{ex9} elucidates combinatorial aspects of
dimensional analysis and may be useful for exploratory research. The
formal notions developed in this section could also serve as a foundation
for computer-assisted dimensional analysis, as sketched in \cite{key-6}.
\begin{rem}
There is an ambivalence in classical dimensional analysis regarding
multiple-equation results. Augmented dimensional analysis yields one
equation for every prebasis for a quantity function. Similarly, in
some formulations of mainly classical dimensional analysis groups
of ``repeating variables'' are distinguished; each such group is
just a prebasis for a quantity function in the present approach. However,
when it has not been overlooked or ignored, the fact that dimensional
analysis may yield more than one possible equation has often been
regarded as a non-uniqueness problem, dealt with by arbitrarily choosing
one equation or by introducing pragmatic rules to single out the ``right''
equation \cite[p. 300]{whi}, \cite[p, 2]{key-11}. By contrast, augmented
dimensional analysis generates a system of simultaneous equations,
all of which can be retained. The guiding principle is that equations
that contain the same information should be merged into one equation,
but if an equation cannot be derived from the other equations (if
any) then it should be retained, since otherwise useful information
could be lost.
\end{rem}

\section{Concluding remarks}

We have seen that dimensional analysis is simply a mathematical technique
that can be used to apply and test certain theories that admit mathematical
formalizations. This technique belongs to a big family of approaches
that are based on the principle of covariance \cite{bar,lo}: a relation
between scalars representing a certain relation between fundamental
objects relative to a reference frame must continue to hold when the
frame is changed to another one of the same kind. The scalars may
change, but they change in tandem. In physics, the reference frames
are often inertial frames in an affine space; in dimensional analysis,
the reference frames are bases for a quantity space, and the fundamental
objects are quantities. A ``physically meaningful'' complete quantity
function is precisely one with a covariant scalar representation.
While it is an important principle, underlying classical dimensional
analysis in terms of scalars, that a function which is a covariant
scalar representation of a quantity function must have a special form,
a quantity function which admits a covariant scalar representation
must also have a special form \textendash{} see the proof of Theorem
\ref{pimain} and Examples \ref{ex1} and \ref{x3-1}. This principle
of quantity-measure duality presupposes a clear distinction between
quantities and their scalar representations, elaborated in quantity
calculus \cite{jo3}.

Theorem \ref{pimain} applies to scaling-covariant scalar representations
of quantity functions, not to covariant scalar representations in
general, so the premise that the quantity function considered is ``physically
meaningful'' in the sense of possessing the latter kind of scalar
representation is not fully used. It is not immediately obvious how
representation theorems where covariance under a more general form
of basis change is assumed should be formulated and proved, and if
they would produce stronger results. We leave these questions as open
problems.

Classical dimensional analysis can be formulated without reference
to products of quantities by relying on a $\pi$ theorem expressed
in terms of ``dimensionless'' products $p_{i}$ of measures of quantities;
these are products which are invariant under a scaling of fundamental
units of measurement. In augmented dimensional analysis, the ``dimensionless''
products are instead $\pi$-monomials $\pi_{i}\in\left[1_{Q}\right]$,
products of quantities corresponding to pseudocircuits. This approach
requires more background, but also leads to a deeper understanding
of dimensional analysis. A system with a set of quantities, a basis
of quantities and corresponding measures of quantities is completely
analogous to a system with a set of vectors, a basis of vectors and
corresponding tuples of vector coordinates, or a system with a set
of tensors, a basis of tensors (often vectors) and corresponding arrays
of tensor components. Nowadays, vectors and tensors are seldom identified
or confused with their scalar representations, and a coordinate-free
approach is widely embraced in linear and multilinear algebra. It
may be time to make this transition also in dimensional analysis.

\subsection*{Acknowledgment}

Appendix A is based on an idea and a proof sketch by Álvaro P. Raposo.

\appendix

\section{Solutions of the functional equations $\Psi_{1}\left(x\right)=x\,\Psi_{1}\left(x^{-1}\right)$
and $\Psi_{2}\left(x\right)=x^{-1}\Psi_{2}\left(x^{-1}\right)$ under
analyticity conditions}

Note that a quantity function $\Psi:\left[1_{Q}\right]\rightarrow\left[1_{Q}\right]$
where $Q$ is a quantity space over $\mathbb{R}$ can be identified
with a scalar function $\psi:\mathbb{R}\rightarrow\mathbb{R}$, so
we can replace the quantity functions $\Psi_{1}$ and $\Psi_{2}$
in the functional equations $\Psi_{1}\left(x\right)=x\,\Psi_{1}\left(x^{-1}\right)$
and $\Psi_{2}\left(x\right)=x^{-1}\Psi_{2}\left(x^{-1}\right)$ by
the real functions $\psi_{1}$ and $\psi_{2}$.

First consider the equation 
\begin{equation}
\psi_{1}\left(x\right)=x\,\psi_{1}\left(x^{-1}\right)\qquad\left(x\neq0\right).\label{eq:001-1}
\end{equation}
Assume that $\psi_{1}$ is infinitely differentiable at $x=0$, and
that the Taylor series about $0$ converges to $\psi_{1}$ on $\mathbb{R}$.
Also assume that the corresponding complex Taylor series exists as
well and converges on $\mathbb{C}$ to a complex function $\varPsi_{1}$
such that $\varPsi_{1}\left(z\right)=\psi_{1}\left(x\right)$ for
$z=x$.%
{} Then $\varPsi_{1}$ has a unique Taylor series expansion about $0$
of the form
\begin{equation}
\varPsi_{1}\left(z\right)=a_{0}+a_{1}z+\sum\nolimits _{n=2}^{\infty}a_{n}z^{n}.\label{eq:02-1}
\end{equation}
and hence the function $\varPsi_{1}'$ defined by $\varPsi_{1}'\left(z\right)=\varPsi_{1}\left(z^{-1}\right)$
for $z\neq0$ has a unique Laurent series expansion about $0$ of
the form
\begin{equation}
\varPsi_{1}'\left(z\right)=\sum_{n=-\infty}^{\infty}\nolimits a_{n}'z^{n}=a_{0}+a_{1}z^{-1}+\sum_{n=2}^{\infty}\nolimits a_{n}z^{-n}\qquad\left(z\neq0\right),\label{eq:L1}
\end{equation}
so identifying coefficients we have $a_{n}'=0$ for all $n\geq1$.
From (\foreignlanguage{british}{\ref{eq:001-1}}) we obtain
\[
\varPsi_{1}'\left(z\right)=z^{-1}\varPsi_{1}\left(z\right)\qquad(z\neq0),
\]
so (\foreignlanguage{british}{\ref{eq:02-1}}) gives another Laurent
series expansion of $\varPsi_{1}^{'}$ about $0$,
\begin{equation}
\varPsi_{1}'\left(z\right)=z^{-1}\varPsi_{1}\left(a_{0}+a_{1}z+\sum_{n=2}^{\infty}\nolimits a_{n}z^{n}\right)=a_{0}z^{-1}+a_{1}+\sum_{n=2}^{\infty}\nolimits a_{n}z^{n-1}\qquad(z\neq0)\label{eq:L2}
\end{equation}

Comparing (\foreignlanguage{british}{\ref{eq:L1}}) and (\foreignlanguage{british}{\ref{eq:L2}}),
we conclude that $a_{0}=a_{1}$ and $a_{n+1}=a_{n}'=0$ for all $n\geq1$.
From (\foreignlanguage{british}{\ref{eq:02-1}}) we thus obtain $\varPsi_{1}\left(z\right)=k_{1}\left(1+z\right)$,
so the solution of (\foreignlanguage{british}{\ref{eq:001-1}}) under
given assumptions is
\begin{equation}
\psi_{1}\left(x\right)=k_{1}\left(1+x\right).\label{eq:sol1}
\end{equation}

We also consider the equation 
\begin{equation}
\psi_{2}\left(x\right)=x^{-1}\psi_{2}\left(x^{-1}\right)\qquad\left(x\neq0\right).\label{eq:04-1}
\end{equation}
Let $\psi_{2}'$ be the real function such that 
\[
\psi_{2}'\left(x\right)=\psi_{2}\left(x\right)^{-1}\qquad\left(\psi_{2}\left(x\right)\neq0\right).
\]
Assume that $\psi_{2}'$ and the corresponding complex function $\varPsi{}_{2}'$
satisfy the same kind of conditions as $\psi_{1}$ and $\varPsi_{1}$,
respectively. It follows from (\foreignlanguage{british}{\ref{eq:04-1})
that $\psi_{2}\left(x\right)^{-1}=\left(x^{-1}\right)^{-1}\psi_{2}\left(x^{-1}\right)^{-1}$,
so
\[
\psi_{2}'\left(x\right)=x\,\psi_{2}'\left(x^{-1}\right)\qquad(x,\psi_{2}\left(x\right)\neq0),
\]
 and this equation has the same form as (\ref{eq:001-1}), so in analogy
with (\ref{eq:sol1}) we obtain
\[
\psi_{2}'\left(x\right)=k_{2}\left(1+x\right),
\]
or equivalently }$\psi_{2}\left(x\right)^{-1}=k_{2}\left(1+x\right)$,
so assuming that $k_{2},1+x\neq0$ we have
\begin{equation}
\psi_{2}\left(x\right)=k_{2}^{-1}\left(1+x\right)^{-1}.\label{sol2}
\end{equation}

We have thus shown that the functional equations in Examples \foreignlanguage{british}{\ref{x4}},
\foreignlanguage{british}{\ref{x5}} and \foreignlanguage{british}{\ref{x6}}
have solutions unique up to a constant of proportionality under suitable
conditions that mainly concern analyticity.

Conversely, it should be noted that there are functions which satisfy
one of the functional equations, but do not satisfy the corresponding
analyticity conditions and also do not have the form (\foreignlanguage{british}{\ref{eq:sol1})
or (\ref{sol2})}. For example, if $\psi_{1}\left(x\right)=\sqrt{1+x^{2}}$
then $\psi_{1}$ satisfies the equation $\psi_{1}\left(x\right)=x\psi_{1}\left(x^{-1}\right)$,
but it can be shown that the Taylor series $\sum_{n=0}^{\infty}\left(\psi_{1}^{(n)}\left(0\right)/n!\right)x^{n}$
does not converge to $\psi_{1}$ on $\mathbb{R}$.

Specifically, the Taylor series for the complex function $\varPsi_{1}$
converges only for $\left|z\right|<r\leq1$ since the derivative
\[
\frac{d}{dz}\varPsi_{1}\left(z\right)=\frac{z}{\sqrt{1+z^{2}}}
\]
is not defined for $z=\pm i$, and the distance from $0$ to these
singularities in the complex plane is $1$. Thus. the Laurent series
for $z^{-1}\varPsi_{1}$ converges for $0<\left|z\right|<r\leq1$,
whereas the Laurent series for $\varPsi_{1}'$ converges for $\left|z\right|>1/r\geq1$.
This means that the two Laurent series are defined on non-overlapping
domains, so the argument from the non-uniqueness of the Laurent series
expansion does not apply in this case.%


\begin{thebibliography}{10}
\bibitem{bo}de Boer, J. (1994). On the history of quantity calculus
and the international system, \emph{Metrologia}, \textbf{31} 405\textendash 429.
https://doi.org/10.1088/0026-1394/31/6/001.

\bibitem{bra}Brand, L. (1957). The Pi Theorem of Dimensional Analysis.
\emph{Arch Rational Mech Anal}, \textbf{1} 35-45. https://doi.org/10.1007/BF00297994.

\bibitem{bar}Barenblatt, G.I. (1996). \emph{Scaling, self-similarity,
and intermediate asymptotics}. Cambridge University Press.

\bibitem{bri}Bridgman, P.W. (1922). \emph{Dimensional Analysis}.
Yale University Press.

\bibitem{bu}Buckingham, E. (1914). On physically similar systems:
illustrations of the use of dimensional equations\emph{. Phys Rev},
\textbf{4} 345\textendash 376.

\bibitem{ca}Carlson, D.E. (1978). A mathematical theory of physical
units, dimensions and measures. \emph{Arch Rational Mech Anal}, \textbf{70}
289\textendash 304. https://doi.org/10.1007/BF00281156.

\bibitem{dro}Drobot, S. (1953). On the foundations of dimensional
analysis. \emph{Studia Mathematica}, \textbf{14} 84\textendash 99.

\bibitem{fe}Federman, A. (1911). On some general methods of integration
of first-order partial differential equations. \emph{Proceedings of
the Saint-Petersburg polytechnic institute. Section of technology,
natural science and mathematics}, \textbf{16} 97\textendash 155. (In
Russian.)

\bibitem{fle}Fleischmann, R. (1951). Die Struktur des physikalischen
Begriffssystems. \emph{Z. Physik} \textbf{129} 377\textendash 400.
https://doi.org/10.1007/BF01379590.

\bibitem{fo}Fourier, J. (1822). \emph{Théorie analytique de la chaleur}.
Paris. 

\bibitem{gi}Gibbings, J.C. (2011). \emph{Dimensional Analysis}. Springer.

\bibitem{har}Hardtke, J-D. (2019). On Buckingham's $\Pi$-theorem.
arXiv:1912.08744 {[}math-ph{]}. https://doi.org/10.48550/arXiv.1912.08744.

\bibitem{key-1}Janyška, J., Modugno, M., Vitolo, R. (2010). An Algebraic
Approach to Physical Scales, \emph{Acta Appl Math} \textbf{110} 1249\textendash 1276.
https://doi.org/10.1007/s10440-009-9505-6.

\bibitem{jo1}Jonsson, D. (2014). Quantities, Dimensions and Dimensional
Analysis. arXiv:1408.5024{[}math.HO{]}. https://doi.org/10.48550/arXiv.1408.5024.

\bibitem{key-6}Jonsson, D. (2014). Dimensional Analysis: A Centenary
Update. arXiv:1411.2798{[}math.HO{]}. https://doi.org/10.48550/arXiv.1411.2798.

\bibitem{jo2}Jonsson, D. (2020). An Algebraic Foundation of Amended
Dimensional Analysis. arXiv:2010.15769 {[}math-ph{]}. https://doi.org/10.48550/arXiv.2010.15769.

\bibitem{jo3}Jonsson, D. (2023). Scalable monoids and quantity calculus.
\emph{Semigroup Forum} \textbf{107} 158\textendash 187. https://doi.org/10.1007/s00233-023-10371-0.

\bibitem{key-2}Kock, A. (1989). Mathematical structure of physical
quantities. \emph{Arch Rational Mech Anal} \textbf{107} 99-104. https://doi.org/10.1007/BF00286495.

\bibitem{la}Langhaar, H.L. (1951). \emph{Dimensional Analysis and
Theory of Models}. Wiley.

\bibitem{lo}Longo, S.G. (2021). \emph{Principles and Applications
of Dimensional Analysis and Similarity}, Mathematical Engineering,
https://doi.org/10.1007/978-3-030-79217-6. Springer.

\bibitem{key-11} (2021) Mokbel, K., Saad, T. BuckinghamPy: A Python
software for dimensional analysis, \emph{SoftwareX}, \textbf{16} 100851,
https://doi.org/10.1016/j.softx.2021.100851.

\bibitem{qua}Quade, W. (1961). Über die algebraische Struktur des
Größenkalküls der Physik. \emph{Abhandlungen der Braunschweigischen
Wissenschaftlichen Gesellschaft}, \textbf{13}, 24\textendash 65.

\bibitem{ra1}Raposo, A.P. (2018). The Algebraic Structure of Quantity
Calculus. \emph{Meas Sci Rev}, \textbf{18} 147-157. https://doi.org/10.1515/msr-2017-0021.

\bibitem{ra2}Raposo, A.P. (2019). The Algebraic Structure of Quantity
Calculus II: Dimensional Analysis and Differential and Integral Calculus.
\emph{Meas Sci Rev}, \textbf{19} 70\textendash 78. https://doi.org/10.2478/msr-2019-0012.

\bibitem{se}Sedov, L.I. (1993). \emph{Similarity and Dimensional
Methods in Mechanics}, 10th ed. CRC Press, Boca Raton.

\bibitem{sze}Szekeres, P. (1978). The Mathematical Foundations of
Dimensional Analysis and the Question of Fundamental Units. \emph{Int
J Theor Phys}, \textbf{17}, 957\textendash 974. https://doi.org/10.1007/BF00678423.

\bibitem{va}Vaschy, A. (1892). Sur les lois de similitude en physique.
\emph{Annales Télégraphiques}, \textbf{19} 25\textendash 28.

\bibitem{whi}White, F.M. (2011).\emph{ Fluid Mechanics}, 7th ed.
McGrawHill, New York.

\bibitem{whi-1}Whitney, H. (1968). The mathematics of physical quantities:
Part II: Quantity structures and dimensional analysis. \emph{Am Math
Mon}, \textbf{75} 227\textendash{} 256. https://doi.org/10.1080/00029890.1968.11970972.
\end{thebibliography}
\end{document}